\documentclass[trackchanges,twocolumn]{aastex701}
\usepackage{amsmath}
\usepackage{gensymb}

\begin{document}

\title{The Binary Properties of Stellar Streams are Set by Cluster Dynamics}

\author[orcid=0009-0005-1914-974X,sname='Phillips']{Anya Phillips}
\affiliation{Center for Astrophysics $|$ Harvard \& Smithsonian, 60 Garden St, Cambridge, MA 02138, USA}
\email[show]{anya.phillips@cfa.harvard.edu}  

\author[orcid=0000-0002-1590-8551]{Charlie Conroy}
\affiliation{Center for Astrophysics $|$ Harvard \& Smithsonian, 60 Garden St, Cambridge, MA 02138, USA}
\email[]{cconroy@cfa.harvard.edu}  

\author[orcid=0000-0001-8042-5794]{Jacob Nibauer}
\affiliation{Department of Astrophysical Sciences, Princeton University, 4 Ivy Lane, Princeton, NJ 08544, USA}
\email{jnibauer@princeton.edu}

\author[orcid=0000-0001-8713-0366]{Long Wang}
\affiliation{School of Physics and Astronomy, Sun Yat-sen University, Daxue Road, Zhuhai 519082, People’s Republic of China}
\email{longwang.astro@live.com}

\author[orcid=0000-0002-0572-8012]{Vedant Chandra}
\affiliation{Center for Astrophysics $|$ Harvard \& Smithsonian, 60 Garden St, Cambridge, MA 02138, USA}
\email{vedant.chandra@cfa.harvard.edu}

\author[orcid=0000-0002-7846-9787]{Ana Bonaca}
\affiliation{The Observatories of the Carnegie Institution for Science, 813 Santa Barbara Street, Pasadena, CA 91101, USA}
\email{abonaca@carnegiescience.edu}

\author[0000-0002-1468-9668]{Jay Strader}
\affiliation{Center for Data Intensive and Time Domain Astronomy, Department of Physics and Astronomy, Michigan State University, East Lansing, MI 48824, USA}
\email{straderj@msu.edu}

\author[orcid=0000-0002-1417-8024]{Morgan MacLeod}
\affiliation{Center for Astrophysics $|$ Harvard \& Smithsonian, 60 Garden St, Cambridge, MA 02138, USA}
\email{morgan.macleod@cfa.harvard.edu}


\begin{abstract}
We present a suite of direct $N$-body simulations of low mass ($<10^4~M_{\odot}$) globular cluster streams initialized with observationally-motivated binary demographics in order to understand the effect of in-cluster dynamical processing on the stream binary population. The models are initialized with a range of stellar densities and cluster orbits, and Poisson variation in the number of massive and short-lived stars. 
Wide binaries are disrupted on short timescales by internal tides and on long timescales by two-body encounters. Tides are most important prior to impulsive mass loss-driven cluster expansion. 
Close binaries ($P_{\rm orb}<10^2~\rm yr$) are most abundant at the stream center due to cluster mass segregation. 
The wide binary fraction and the degree of binary segregation in the resulting stream are sensitive to the initial cluster density and massive star fraction. 
In mock radial velocity surveys of the simulated streams, undetectable binaries have velocity amplitudes of $\sim$$0.5$--$1~\rm km~s^{-1}$, adding $\sim0.1~\rm km\ s^{-1}$ of velocity dispersion to the streams, and are dynamically depleted by $\sim10$--$60\%$ compared to the initial binary population. 
Custom $N$-body models of Milky Way streams with binaries will allow a holistic understanding of their dynamical structures in advance of upcoming multi-epoch spectroscopic surveys. 

\end{abstract}

\keywords{\uat{Stellar streams}{2166} --- \uat{Globular star clusters}{656} --- \uat{Binary stars}{154}} 


\section{Introduction}\label{sec:1}

The stellar streams of disrupting globular clusters (GCs) are sensitive gravitational probes of the Galaxy. They form as a cluster undergoes dynamical relaxation and escaping stars are lost through small apertures around its Lagrange points, creating kinematically cold tidal tails with velocity dispersions of order $1$ km s$^{-1}$ \citep{baumgardtMeanProperMotions2019, giallucaVelocityDispersionGD12021, bonaca2025}. In addition to being dynamically cold, streams possess relatively smooth surface densities, save for predictable dynamical effects such as epicycles \citep{kupper2008-epicycles, kupper2012-streaklines, fardal2015, amoriscoFeathersBifurcationsShells2015}. Large scale variations in the Galactic potential such as the bar, Magellanic clouds, and disk tilting also impart perturbations to stream kinematics and morphology \citep{Hattori2016,bonacaVariationsWidthDensity2020,erkalTotalMassLarge2019,nibauerSlantFanNarrow2024}. 

Impulsive impacts from passing small-scale substructures impart characteristic velocity signatures across the perturbed regions of streams, creating gaps, density variations, and bifurcations. It is possible to infer the perturber's mass, size, and impact geometry given sufficient spatial and kinematic information about the stream stars \citep{johnstonHowLumpyMilky2002,ibataUncoveringColdDark2002, carlbergStarStreamFolding2009, yoon2011,erkal2015a,erkal2015b,sandersDynamicsStreamsubhaloInteractions2016, bovyLinearPerturbationTheory2017,bonacaSpurGapGD12019,hilmiInferringDarkMatter2024, luDetectabilityDarkMatter2025}. Moreover, accumulated perturbations act to heat the streams, so that properties of the population of perturbers may in principle be inferred statistically from the stream velocity dispersion \citep{carlbergDarkMatterSubhalo2012, banikEffectsBaryonicDark2019, banikEvidencePopulationDark2021,Nibauer2025_measurement}. 

The existence and abundance of stream perturbations from small dark matter subhalos is sensitive to the nature of dark matter itself. For example, weakly-interacting massive particle dark matter models predict subhalos down to Earth mass \citep{greenFirstWIMPyHalos2005, wangUniversalStructureDark2020a}, and primordial black holes are predicted to have masses comparable to asteroids ($\approx 10^{-15}~M_{\odot}$, \citealt{carrPrimordialBlackHoles2020}). On the other hand, in ultralight axion models, dark matter substructure is suppressed below $<10^{10} M_{\odot}$ \citep{huFuzzyColdDark2000,huiUltralightScalarsCosmological2017}. Below $\sim10^7 M_{\odot}$, dark matter subhalos are expected to contain no baryonic matter \citep{benitez-llambayDetailedStructureOnset2020,nadlerMilkyWaySatellite2020, nadlerImpactMolecularHydrogen2025}. Instead, the halos most sensitive to the underlying dark matter physics may be detected via their gravitational influence on luminous systems.

Binary stars are a known source of bias in dynamical studies of stellar systems because their individual orbital motions artificially inflate the measured velocity dispersion of the system (e.g., \citealt{kouwenhoven2008,gieles2010,Pianta2022, wirth2024}). In GC streams, added dispersion from binaries could obscure the subtle velocity signatures of subhalo impacts or be mistaken for heating from subhalos. 
Close binaries can be identified over multiple observations based on radial velocity (RV) variability (e.g., \citealt{badenesStellarMultiplicityMeets2018b, Nibauer2025_measurement}) and accounted for in further analysis, but wide binaries with long orbital periods elude detection even with decade-long RV monitoring. 

The abundance of wide binaries in part reflects the density of their formation environments because dense environments tend to disrupt them \citep{scallyWideBinariesOrion1999, parkerBinariesClustersForm2009, gellerSearchThermalEccentricity2019, deaconWideBinariesAre2020}. 
For example, the high density of low-metallicity star formation early in the Milky Way's evolution may explain the observed decrease in the wide binary fraction to low [Fe/H] \citep{hwangWideBinariesH32022}. 
However, whether wide (and undetectable) binaries typically escape their progenitor clusters and enter stellar streams as bound systems is not well known. \citet{roznerBornBeWide2023} analytically predict the demographics of wide binaries that survive cluster dissolution, showing that the overall wide binary fraction scales inversely to the cluster mass. Recent $N$-body calculations suggest a binding energy-dependent binary survival probability, where binaries with orbital periods $\gtrsim 10^4~\rm yr$ are strongly depleted in the early evolution of simulated clusters embedded in a gas potential \citep{wuStatisticalStudyBinary2026}. Such systems have been shown to survive in simulations provided early escape from the cluster into the forming stream \citep{Wang2024}. This result is supported by limited observational evidence that the formation of streams may allow the survival of binaries that would otherwise be destroyed within the clusters. \citet{usmanMultiplePopulationsCH2024} report the discovery of an $s$-process element-rich CH star (which is speculated to be a product of binary mass transfer) in the 300S GC stream, despite the rarity of such systems present within GCs. \citet{sharmaTidalTailsOpen2025} report photometric binary fractions in the tidal tails of open clusters which exceed the binary fractions within the clusters themselves. 
Finally, simulation-based studies of tidal tails reveal them to be potential formation environments for binaries with orbital separations $\gtrsim0.1$ pc \citep{penarrubia2021}. 

Much like stellar streams, wide binaries are of interest due to their potential to probe the small scale mass distribution of the Galaxy beyond their natal environments. They are sensitive to disruption by passing stars, molecular clouds, compact objects, and the tide of their host's potential \citep{rettererWideBinariesSolar1982, bahcallMaximumMassObjects1985, weinbergDynamicalFateWide1987, jiangEvolutionWideBinary2010, livernoisEvolutionBinaryStars2023,yooEndMACHOEra2004a,chanameDiskHaloWide2004,quinnReportedDeathMACHO2009,carrPrimordialBlackHoles2020}, and may even probe the nature of dark substructures in the Milky Way and nearby dwarf galaxies 
\citep{penarrubiaWideBinariesUltrafaint2016,Ramiriz2023,shariat2025}. Nonetheless, the same dynamical fragility that makes wide binaries interesting gravitational probes also makes their survival in dense cluster environments and presence in streams highly uncertain. 

In this work, we explore the influence of in-cluster dynamics on the binary populations of GC streams through a suite of direct $N$-body models initialized with observationally-motivated binary populations. 
We motivate our choice of simulation grid with theoretical background and make predictions for the behavior of binaries in clusters and streams in Section \ref{sec:theory}. The implementation and analysis of our simulations are detailed in Section \ref{sec:methods}. We present our results in Section \ref{sec:results}, discuss their implications and limitations in Section \ref{sec:discussion}, and conclude in Section \ref{sec:summary}.

\section{Theoretical background}\label{sec:theory}

In this section, we introduce the relationships between key cluster properties, timescales, and the relevant dynamical processes that shape the binary populations of the cluster and the stream. 

The structural evolution of a GC and the formation of its stream occur on three key timescales---the cluster dynamical time, relaxation time, and dissolution time. These in turn are determined by the cluster density, mass, mass loss history, and the external tidal field. 
The dynamical timescale within a cluster is set by the local density $\rho$, 
\begin{equation}\label{eq:t_dyn}
    t_{\rm dyn} \sim \frac{1}{\sqrt{G\rho}},
\end{equation}
and represents the time over which the cluster responds to impulsive dynamical perturbations. For the low-mass clusters studied in this work, typical dynamical timescales at the cluster half mass radius range from $\lesssim1$ to several Myr. Because it is approximately equal to the crossing time $t_{\rm cr}$, the dynamical time determines the two-body relaxation timescale:
\begin{equation}
    \label{eq:t_relax}
    t_{\rm{relax}} \approx \frac{0.1N}{\ln{\Lambda}} t_{\rm{cr}}\sim \frac{0.1 N}{\ln{N}}t_{\rm dyn},
\end{equation}
for a cluster of $N$ stars, where $\ln{\Lambda}\approx\ln{N}$ is the Coulomb logarithm (which captures the range of impact parameters between gravitationally interacting stars). At the half mass radii of clusters studied in this work, their relaxation timescales range from tens of Myr to $\sim1$ Gyr. 
Two-body interactions change the velocity of a star in the cluster by order the initial velocity over $t_{\rm relax}$ (e.g., \citealt{binneyTremaine} $\S 1.2.1$). In this process, a subset of stars acquire speeds in excess of the cluster escape velocity and enter its growing stream. Relaxation continues until the cluster has completely dissolved, at $t_{\rm dis}$, the dissolution time.

During their formation, clusters virialize through violent relaxation (e.g., \citealt{binneyTremaine}, $\S4.10.2$), at the end of which all stars are left with the same energy per unit mass. Two-body relaxation then allows the stars to exchange kinetic energy and approach equipartition (though most clusters do not reach equipartition, even after a number of relaxation timescales). Over $t_{\rm relax}$, massive stars lose kinetic energy and sink to the cluster center (``mass segregation"), while low mass stars move to the cluster outskirts and preferentially escape. 
The degree of cluster mass segregation is therefore determined by the relative lengths of the relaxation timescale and the dissolution time \citep{lamersEvolutionGlobalStellar2013}. 
Clusters with $t_{\rm relax}\ll t_{\rm dis}$ thoroughly mass segregate before their dissolution, but if $t_{\rm dis}\lesssim t_{\rm relax}$, then the cluster dissolves before significant mass segregation can occur. 
In this context, binary stars act as single, more massive particles, and build up in the cores of mass segregated clusters, while single stars preferentially escape into the stream. This behavior has been observed in simulations \citep{hurley2007, fregeauEVOLUTIONBINARYFRACTION2009,weatherford2023}, and in Milky Way GCs, which have binary fractions that decrease with increasing radius (e.g., \citealt{miloneACSSurveyGalactic2012}). 

In addition to mass segregation, the tendency toward energy equipartition in a cluster influences the orbits of binaries. Cluster binaries can be classified based on their orbital separation as ``hard" and ``soft" using the boundary 
\begin{equation}
    \label{eq:ahs}
    a_{\rm{h/s}} = \frac{Gm_1m_2}{\langle m\rangle \sigma^2},
\end{equation}
for binary companion masses $m_1$ and $m_2$, locally-averaged stellar mass $\langle m\rangle$, and local velocity dispersion $\sigma$ (\citealt{heggie1975, hills1975, binneyTremaine}, $\S7.5.7$). This expression is the result of equating the orbital energy of the binary with the average kinetic energy of the surrounding stars. 
Hard binaries, with semimajor axes $a<a_{\rm{h/s}}$, have large binding energies compared to the local velocity dispersion, whereas soft binaries with $a>a_{\rm{h/s}}$ have comparatively small binding energies. As a result, gravitational encounters with typical nearby stars will tend to harden hard binaries and to soften and eventually disrupt soft binaries. For the cluster densities studied in this work, typical hard/soft boundaries lie at orbital periods $\gtrsim10^2~\rm yr$ (although the local velocity dispersion changes this value throughout the cluster's profile). 

Wide binaries may also be disrupted within the cluster due to internal tidal forces. To find the critical binary orbital separation at which this is possible, we follow a derivation similar to that of the Roche limit (see \citealt{Ryden2021}, $\S4.3.1$). We assume that the vector connecting the initial positions of the two stars in a binary of semimajor axis $a$ points radially toward the center of the cluster, and that the two stars are separated by a distance $2a$. Then, if the binary is a distance $r$ from the cluster center, the critical orbital separation at which it can be disrupted by the differential gravity of the enclosed mass within $r$, $M(<r)$, is 
\begin{equation}
    \label{eq:roche_limit}
    a_{\rm{crit}}(r) = \left(\frac{r^3 m_1 m_2}{16 M(<r) (m_1+m_2)}\right)^{1/3}\propto \overline{\rho}^{-1/3},
\end{equation}
where $\overline{\rho}=M(<r)\left(\frac{4\pi}{3}r^3\right)^{-1}$ is an averaged density interior to $r$. In the densest central regions of clusters studied in this work, the orbital periods of binaries that may be disrupted by internal tides are $\gtrsim10^4~\rm yr$ (see Figure \ref{fig:roche_hs}).

Density is the central determinant of the relevant timescales and dynamical processes in GCs, appearing explicitly in the expressions for $t_{\rm dyn}$, $t_{\rm relax}$ and $a_{\rm crit}$, and implicitly in the expression for $a_{\rm h/s}$ (since in virial equilibrium, the velocity dispersion $\sigma$ increases with density). 
Dense clusters adjust to perturbations, undergo two-body relaxation, and mass segregate on shorter timescales, while disrupting more binaries from a given population than diffuse clusters. 
However, it is important to recognize that the density evolves significantly over the cluster lifetime.

\begin{figure}
    \centering
    \includegraphics[width=\linewidth]{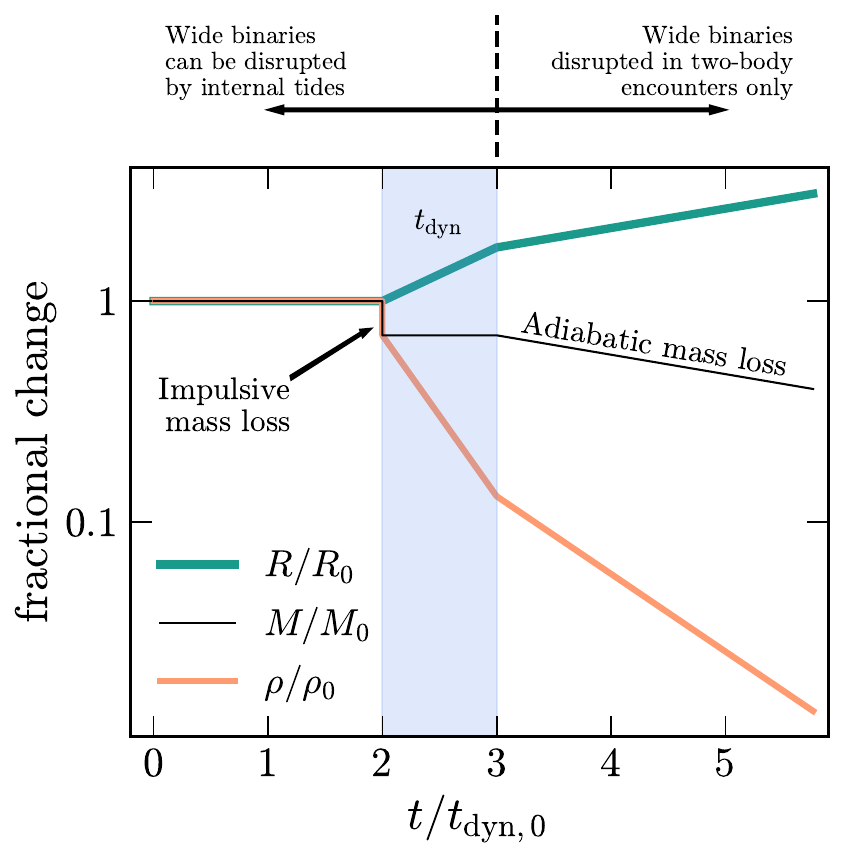}
    \caption{Schematic demonstration of the effects of cluster mass loss on its structure. The black line shows the cluster mass loss history, including impulsive mass loss (which occurs at $t=2~t_{\rm dyn}$ in this example) and adiabatic mass loss (which begins at $t=3~t_{\rm dyn}$). After impulsive mass loss, the cluster radius increases and its density decreases on a dynamical time. During adiabatic mass loss, the radius and density change more gradually. The transition points depend on the stellar population and dynamical state of the cluster. The annotations at the top of the panel note relevant disruption mechanisms for wide binaries in the low mass clusters studied in this work.}
    \label{fig:massloss_demo}
\end{figure}
In addition to processes noted above, clusters lose mass via the winds and deaths of their stars. Cluster mass loss may be ``impulsive," occurring over timescales shorter than the cluster dynamical time, or ``adiabatic," occurring over longer timescales. 
To maintain dynamical equilibrium during mass loss, clusters expand in size and decrease in density on a dynamical time. Figure \ref{fig:massloss_demo} illustrates the effect of impulsive and adiabatic mass loss on the cluster radius and density, with expansion factors calculated according to the analytic description in \citet{hillsEffectMassLoss1980}. For initial cluster mass $M_0$ and size $R_0$ (e.g., the half-mass radius), the mass loss factor $\Delta M/M_0$ causes the expansion factor
\begin{equation}
    \frac{R}{R_0} = \begin{cases}
        \frac{1-\Delta M/M_0}{1-2\Delta M/M_0} & \text{(Impulsive)} \\
        \frac{1}{1-\Delta M/M_0} & \text{(Adiabatic)}
    \end{cases}
\end{equation}
over $t_{\rm dyn}$. Impulsive mass loss causes a larger expansion factor than adiabatic mass loss. 

The winds and supernovae of massive ($\gtrsim 8~M_{\odot}$) stars are the main source of impulsive mass loss, so impulsive mass loss ends with the deaths of these stars at $t\leq$ several tens of Myr \citep{lamersUnderstandingStellarEvolution2017}. Poisson variation in the number of massive stars across clusters with otherwise identical initial conditions can lead to different expansion factors in their early evolution (owing to resulting variations in the degree of rapid wind mass loss) and variation in their long-term dynamical behavior (due to differing levels of binary black hole heating; \citealt{wangImpactMassiveStars2021}). 
Winds of lower-mass stars and the dynamical evaporation of stars from the cluster continue to drive adiabatic mass loss throughout its evolution. 

Cluster expansion increases $t_{\rm dyn}$ and $t_{\rm relax}$ (equations \ref{eq:t_dyn}, \ref{eq:t_relax}), and can unbind the cluster over a shorter dissolution time \citep{hillsEffectMassLoss1980,baumgardtComprehensiveSetSimulations2007}. 
Expansion also mediates the dynamical processing of binaries. The decreased density slows the timescale for mass and binary segregation, and increases the hard/soft boundary. 
The critical orbital separation for binary disruption by internal tides also increases. As an example, for a binary composed of two $0.5~M_{\odot}$ companions separated by $a=2000$ AU (with an orbital period of $\sim10^6$ yr) located at the half-mass radius of a $10^4~M_{\odot}$ host cluster (i.e., so that $M~(<r)\sim5000~M_{\odot}$), the cluster must have a half mass radius $\lesssim0.7~\rm pc$ in order to disrupt the binary with internal tides (Equation \ref{eq:roche_limit}). The cluster density in this example is similar to the densest clusters we include in this work (see Section \ref{sec:methods_1_ICs}). We therefore expect binary disruption by internal tides to be an important disruption mechanism only before the cluster expands significantly due to impulsive mass loss from stellar evolution. This is indicated in the annotations above Figure \ref{fig:massloss_demo}. 

The dynamical effect of mass loss from stellar evolution motivates our definition of a final relevant timescale for binary processing in GCs. The approximate duration of early cluster evolution over which wide binaries may be disrupted by internal tides is given by:
\begin{equation}\label{eq:t_internaltides}
    t_{\rm int\ tid} = t_{\rm SE} + t_{\rm dyn}(t_{\rm SE}),
\end{equation}
where $t_{\rm SE}$ is a characteristic stellar evolution timescale for stars contributing significantly to impulsive mass loss, and $t_{\rm dyn}(t_{\rm SE})$ is the dynamical time at $t_{\rm SE}$, over which the cluster expands due to impulsive mass loss from stellar evolution. Going forward, we take $t_{\rm SE}$ to be the main sequence lifetime of a $25~M_{\odot}$ star, or $\sim6.5$ Myr \citep{lamersUnderstandingStellarEvolution2017}. We choose this stellar mass because the resulting $t_{\rm int\ tid}$ empirically corresponds with the disappearance of all tidally-disruptable binaries in simulated clusters (see Section \ref{sec:results_2_stellarpops} and Figure \ref{fig:depleting_longperiods_late}). 

\begin{deluxetable*}{lllrrrrrrrrl}
    \tablecaption{A summary of simulations used in this work. We include the progenitor orbit shape, initial virial and half mass radii ($R_{\rm vir, 0}$ and $R_{h,0}$), number and total mass of OB stars (with $M\geq 8 M_{\odot}$; $N_{\rm OB}$ and $M_{\rm OB}$), total initial cluster mass ($M_{\rm clust, 0}$), initial dynamical and relaxation timescales at the half mass radius ($t_{\rm dyn, 0}$ and $t_{\rm rh, 0}$), initial 1-dimensional velocity dispersion ($\sigma_{\rm 1D, 0}$), dissolution time ($t_{\rm dis}$), and the time of the progenitor's last pericenter passage before dissolution ($t_{\rm peri}$). For the four clusters with repeat simulations, we include the $16^{\rm th}$--$84^{\rm th}$ percentile ranges in their dissolution times ($\sigma_{t_{\rm dis}}$). The simulations with altered binary fractions are listed in the bottom rows. \label{tab:simulation_grid}}
    \tablehead{ 
      \colhead{Orbit} & \colhead{$R_{\rm{vir},0}$} & \colhead{$R_{h,0}$} & \colhead{$N_{\rm{OB}}$} & \colhead{$M_{\rm{OB}}$} & \colhead{$M_{\rm{clust},0}$} & \colhead{$t_{\rm{dyn},0}$} & \colhead{$t_{{rh},0}$} & \colhead{$\sigma_{\rm{1D},0}$} & \colhead{$t_{\rm{dis}}$} & \colhead{$t_{\rm{peri}}$} & $\sigma_{t_{\rm dis}}$
     \\
     \colhead{} & \colhead{pc} &\colhead{pc}& \colhead{} & \colhead{$M_{\odot}$} & \colhead{$M_{\odot}$} & \colhead{Myr}& \colhead{Myr} & \colhead{km s$^{-1}$} & \colhead{Myr} & \colhead{Myr} & Myr 
     }
    \startdata
        \multicolumn{12}{l}{\textbf{Main grid}:}  \\ [1mm]
        Circular & 0.75 & 0.63 & 86 & 1621 & 8499 & 0.4 & 66 & 1.46 & 13388 & 13388  &  \\
        Circular & 0.75 & 0.64 & 137 & 2915 & 9940 & 0.4 & 61 & 1.36 & 12893 & 12893  &  \\
        Circular & 1.5 & 1.28 & 86 & 1621 & 8499 & 1.0 & 156 & 1.01 & 10784 & 10784  &  \\
        Circular & 1.5 & 1.32 & 137 & 2915 & 9940 & 1.0 & 148 & 0.98 & 7298 & 7298  &  \\
        Circular & 3.0 & 2.36 & 86 & 1621 & 8499 & 2.8 & 431 & 0.81 & 13929 & 13929  &  \\
        Circular & 3.0 & 2.55 & 137 & 2915 & 9940 & 2.7 & 427 & 0.78 & 4885 & 4885  &  \\
        Circular & 6.0 & 4.62 & 86 & 1621 & 8499 & 8.2 & 1273 & 0.57 & 7365 & 7365  &  \\
        Circular & 6.0 & 5.01 & 137 & 2915 & 9940 & 6.7 & 1050 & 0.61 & 3065 & 3065  &  \\
        GD-1 & 0.75 & 0.63 & 86 & 1621 & 8499 & 0.4 & 66 & 1.46 & 10853 & 10499 & 2400 \\
        GD-1 & 0.75 & 0.64 & 137 & 2915 & 9940 & 0.4 & 61 & 1.36 & 5929 & 5552 & 3320 \\
        GD-1 & 1.5 & 1.28 & 86 & 1621 & 8499 & 1.0 & 156 & 1.01 & 11430 & 10990  &  \\
        GD-1 & 1.5 & 1.32 & 137 & 2915 & 9940 & 1.0 & 148 & 0.98 & 4321 & 4081  &  \\
        GD-1 & 3.0 & 2.36 & 86 & 1621 & 8499 & 2.8 & 431 & 0.81 & 10928 & 10488  &  \\
        GD-1 & 3.0 & 2.55 & 137 & 2915 & 9940 & 2.7 & 427 & 0.78 & 3354 & 3096  &  \\
        GD-1 & 6.0 & 4.62 & 86 & 1621 & 8499 & 8.2 & 1273 & 0.57 & 7004 & 6549 & 830 \\
        GD-1 & 6.0 & 5.01 & 137 & 2915 & 9940 & 6.7 & 1050 & 0.61 & 2468 & 2108 & 770 \\
        Pal 5 & 0.75 & 0.63 & 86 & 1621 & 8499 & 0.4 & 66 & 1.46 & 5405 & 5164  &  \\
        Pal 5 & 0.75 & 0.64 & 137 & 2915 & 9940 & 0.4 & 61 & 1.36 & 2399 & 2255  &  \\
        Pal 5 & 1.5 & 1.28 & 86 & 1621 & 8499 & 1.0 & 156 & 1.01 & 5381 & 5169  &  \\
        Pal 5 & 1.5 & 1.32 & 137 & 2915 & 9940 & 1.0 & 148 & 0.98 & 2776 & 2546  &  \\
        Pal 5 & 3.0 & 2.36 & 86 & 1621 & 8499 & 2.8 & 431 & 0.81 & 3956 & 3710  &  \\
        Pal 5 & 3.0 & 2.55 & 137 & 2915 & 9940 & 2.7 & 427 & 0.78 & 1364 & 1090  &  \\
        Pal 5 & 6.0 & 4.62 & 86 & 1621 & 8499 & 8.2 & 1273 & 0.57 & 2341 & 2257  &  \\
        Pal 5 & 6.0 & 5.01 & 137 & 2915 & 9940 & 6.7 & 1050 & 0.61 & 1315 & 1089  &  \\
        \tableline
        \multicolumn{11}{l}{\textbf{With altered binary populations}:} & Notes \\ [1mm]
        GD-1 & 1.5 & 1.28 & 147 & 2895 & 11184 & 0.9 & 138 & 1.37 & 7330 & 7050 & \parbox[t]{2cm}{\footnotesize Doubled binary fraction}\\ 
        GD-1 & 1.5 & 1.27 & 91  & 1711 & 8827  & 1.0 & 153 & 1.05 & 8720 & 8520 & \parbox[t]{2cm}{\footnotesize Doubled binary fraction for $M_1\leq 1.2\,M_{\odot}$ only}\\
    \enddata
\end{deluxetable*}

The formation of stellar streams occurs as a cluster dissolves in an external tidal field. The tidal radius of the cluster, or the position of its inner Lagrange point, can be approximated by the Jacobi radius from the restricted three-body problem
\begin{equation}
    \label{eq:rtid}
    r_{\rm{tid}}\approx\left(\frac{M_{\rm cl}}{fM_{\rm h}(<r_{\rm gal})}\right)^{1/3}r_{\rm gal},
\end{equation}
for a cluster of mass $M_{\rm cl}$ orbiting a host of mass $M_{\rm h}$ at a galactocentric distance $r_{\rm gal}$, and a factor $f$ depending on the host potential \citep{szebehely2012theory,valtonen2006three,bonaca2025}. 
Compact GCs typically maintain a tidal radius larger than their size, and dissolve in the ``relaxation limited" regime, where escaping stars have gradually acquired the escape speed via two-body relaxation and are deposited from the cluster Lagrange points. 
For clusters on eccentric orbits, the tidal radius varies with the radial position of the cluster in the external potential, resulting in an increased rate of dissolution at pericenter \citep{amoriscoFeathersBifurcationsShells2015, fardal2015} and an earlier dissolution time. The dissolution of very diffuse GCs with small pericenters may become ``tidally limited," with the cluster size significantly filling or exceeding its tidal radius. 
Stars released in the tidal rather than the relaxation limited regime undergo comparatively less dynamical processing through two-body relaxation. 

In summary, a complex interplay between early cluster expansion, two-body relaxation, and the evolving tidal radius govern the formation of GC streams and their binary populations. In the following sections, we explore these trends numerically with $N$-body simulations.

\section{Methods}\label{sec:methods}

In this section, we introduce our grid of cluster initial conditions ($\S$\ref{sec:methods_1_ICs}), our implementation of $N$-body simulations to evolve them ($\S$\ref{sec:methods_2_impl}), and our analysis procedure for the simulation outputs ($\S$\ref{sec:methods_3_analysis}).

\subsection{Initial Conditions}\label{sec:methods_1_ICs}

Our simulation grid varies the initial density, the fraction of massive, short-lived stars, and the orbit of the cluster. Each cluster has a fixed number of stars, $N=15000$, drawn from the stellar initial mass function. This yields cluster masses comparable to the lowest-mass GCs, $\lesssim 10^4~M_{\odot}$. We summarize the grid variables in Table \ref{tab:simulation_grid} and provide additional detail below.

The clusters are initialized with \citet{king1966} profiles, with the concentration parameter $W_0=2$, and virial radii of $0.75,\ 1.5,\ 3.0,$ or $6.0$ pc to probe a range of initial densities. Our routine for generating particle phase space positions from the King distribution function is based on the \texttt{king.py} script within the Cluster Monte Carlo package \citep{CMC2022}. We modify the original script to preserve the correct density and velocity dispersion profiles for a multi-mass system with no initial mass segregation. The initial virial radius $R_{\rm vir,0}$ is the size used to scale from \citet{henon1971} $N$-body units to physical units after sampling the King distribution function. Table \ref{tab:simulation_grid} provides both the initial virial radius and initial half mass radius of each cluster. 

\begin{figure}
    \centering
    \includegraphics[width=\linewidth]{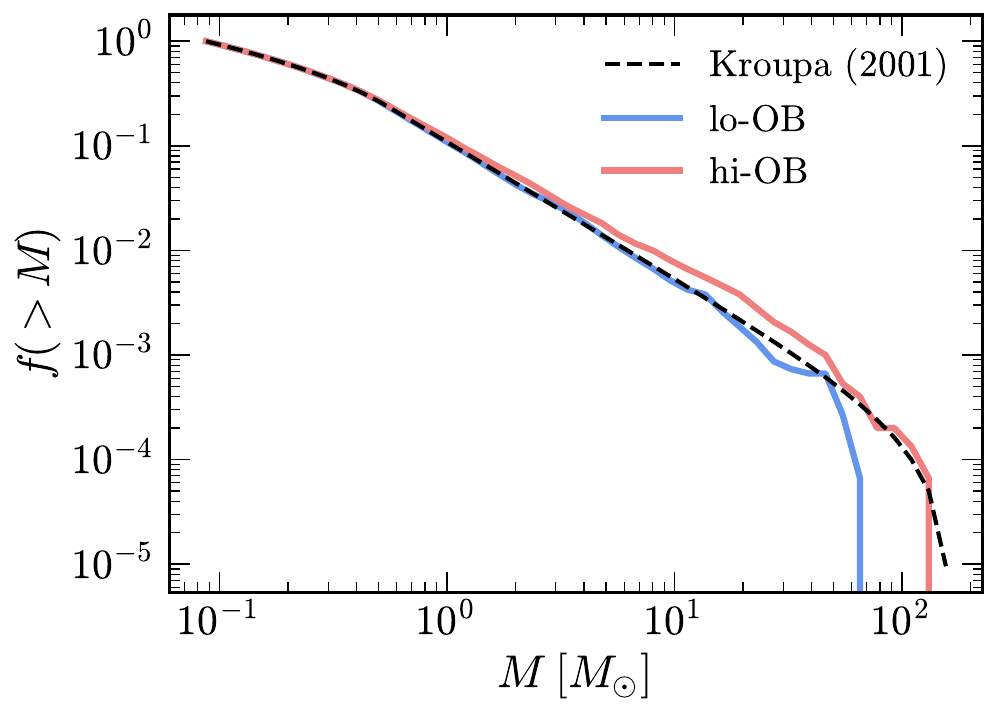}
    \caption{The two samplings of the \citet{kroupa2001} initial mass function used in this work. Poisson variation leads to more high-mass stars in the hi-OB draw than in the lo-OB draw. }
    \label{fig:MF_stochasticity}
\end{figure}

We probe varying degrees of impulsive mass loss by performing two draws of the initial stellar masses from the \citet{kroupa2001} initial mass function (IMF). Poisson variation at high stellar masses leads to variable mass contributions from OB stars: in the first draw, OB stars contribute 1621 $M_{\odot}$ to the cluster (86 stars; going forward, we refer to these clusters as ``lo-OB"), and in the second, they contribute 2916 $M_{\odot}$ (137 stars; ``hi-OB" going forward). We use a definition of OB stars as those with $M \geq 8 M_{\odot}$. We show each draw of the IMF compared to the underlying distribution in Figure \ref{fig:MF_stochasticity}.

We vary the orbit shapes of the progenitor GCs to probe the effect of a time-varying tidal radius on the demographics of binaries that escape into the stream. We choose the orbits of the Milky Way streams GD-1 and Palomar 5 (Pal 5), which both exhibit morphological and velocity substructures and have been studied at length in the context of dark matter constraints (e.g., \citealt{dehnenModelingDisruptionGlobular2004a, bonacaSpurGapGD12019}).
The GD-1 orbit has eccentricity 
\hbox{$e=(r_{\rm{apo}}-r_{\rm{peri}})/(r_{\rm{apo}}+r_{\rm{peri}})=0.33$}, 
with apo- and pericenter  $r_{\rm{apo}}=27.59$ pc and $r_{\rm{peri}}=13.96$ pc. The Pal 5 orbit has $e=0.46$, with
$r_{\rm{apo}}=18.56$ pc and $r_{\rm{peri}}=6.79$ pc. These capture the typical range of orbit sizes for streams---the distribution in $r_{\rm apo}$ for known streams peaks at $\sim 20~\rm kpc$---though we note that observed streams span a much larger range of eccentricities \citep{ibataChartingGalacticAcceleration2024, bonaca2025}. 
We compare streams on GD-1 and Pal 5 orbits to streams on circular orbits of radius $20$ kpc and in the Galactic plane. We emphasize that we do not attempt to produce realistic mocks of the present-day GD-1 and Pal 5 streams, but aim to explore the effect of changing the orbit size and eccentricity on the final binary population.

We compute the appropriate circular velocity in our chosen background potential (see Section \ref{sec:methods_2_impl}) using the package \texttt{gala} \citep{gala} to determine the initial positions of clusters on circular orbits. For the GD-1 and Pal 5 orbits, we back-integrate from the present-day positions of the observed stream progenitors for 10 Gyr using \texttt{gala} to obtain the cluster displacements at $t=0$. We use the present-day progenitor position and velocity for GD-1 estimated by \citet{Nibauer:streamsculptor}, who used data from \citet{starkmanStreamMembersOnly2025}. We use the present-day Pal 5 progenitor ICRS right ascension and declination from \citet{odenkirchenKinematicStudyDisrupting2002}, proper motion from \citet{fritzProperMotionPalomar2015}, and distance and radial velocity from \citet{bovyShapeInnerMilky2016}. 

\begin{figure}
    \centering
    \includegraphics[width=\linewidth]{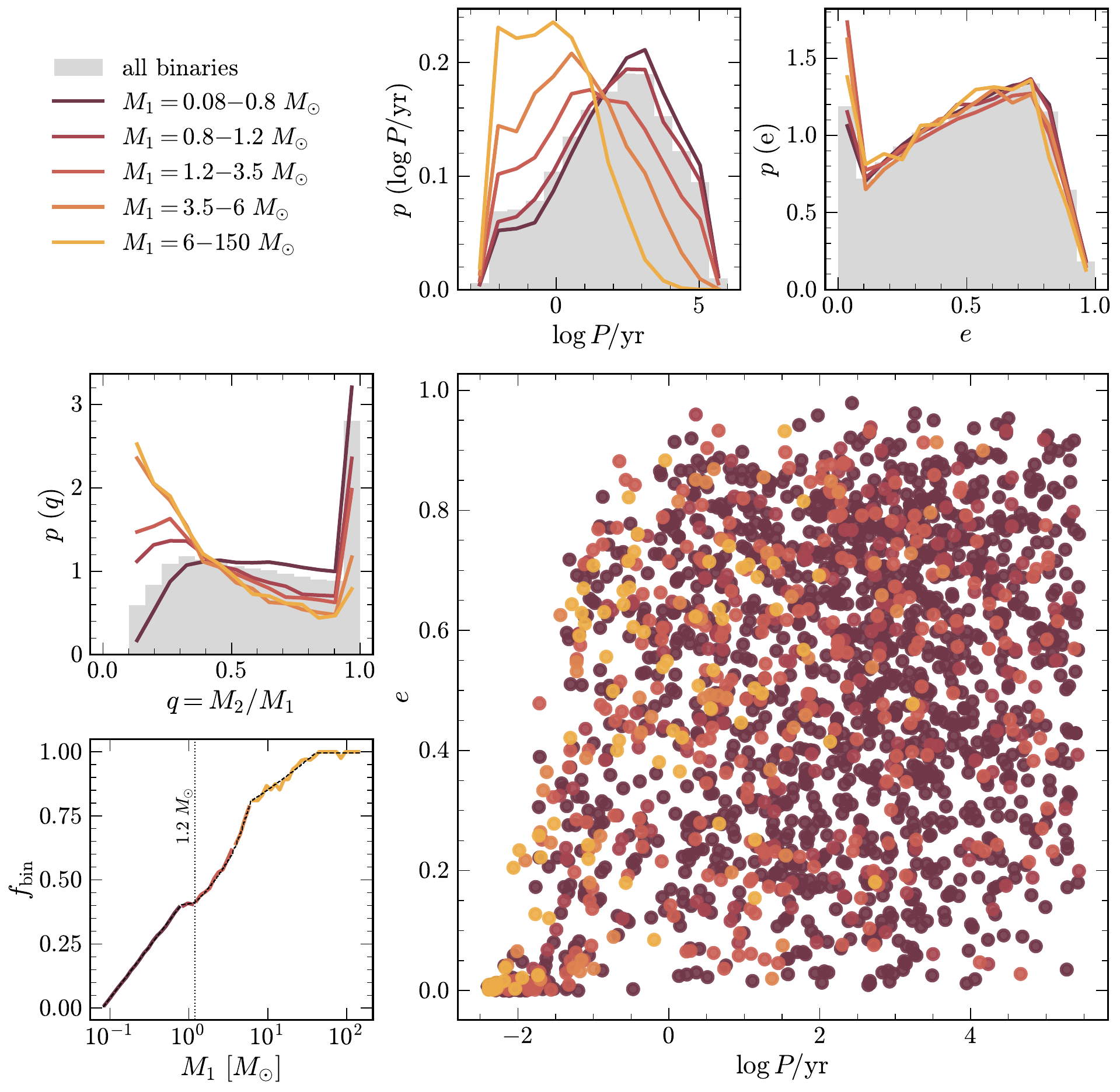}
    \caption{Summary of \texttt{COSMIC} binary demographics used in our initial conditions. We show the distributions in orbital period $P$, eccentricity $e$, and mass ratio $q$ (as probability density functions) and the binary fraction as a function of primary mass for a $N=10^6$ stellar population in small panels. The large panel shows the two-dimensional distribution in eccentricity versus period for a subset of 2000 binaries. We show the total demographic distributions, along with the distributions for the primaries in several mass ranges separately, motivated by the piecewise prescriptions in \citet{moe2017}. The eccentricity distribution includes a peak of circularized systems at short periods, and the mass ratio distribution includes a peak of equal mass systems. The dashed black curve in the bottom left panel shows the underlying binary fraction as a function of primary mass from the \texttt{COSMIC} sampler, and the colored curve shows the resulting sample in bins of primary mass.}
    \label{fig:init_binary_pop}
\end{figure}
We use the binary population synthesis code \texttt{COSMIC} \citep{breivikCOSMICVarianceBinary2020} to generate the stellar population from a multidimensional distribution of binary demographics (binary fractions, periods, eccentricities, and mass ratios). The demographics are constrained observationally for primaries $M_1=0.8\text{--}40~M_{\odot}$ in \citet{moe2017}, and the constraints for the widest systems ($P\gtrsim10^4$~yr) come from binary detections using common proper motion. 

We modify \texttt{COSMIC}'s sampler script \texttt{multidim.py} so that once drawn, the single and binary stars follow a \citet{kroupa2001} initial mass function (as in, for example, \citealt{el-badryGenerativeModelGaia2024}). Henceforth, we will refer to a single star or a binary as a ``system," but note that our cluster size $N=15000$ refers to the total number of \emph{stars}, not the number of systems. 
The total binary fraction of the \texttt{COSMIC} population (that is, $N_{\rm{binaries}}/N_{\rm{systems}}$) is $f_{\rm bin}\sim21\%$. We generate two draws of the \texttt{COSMIC} stellar population in order to obtain the two different cluster mass contributions from OB stars described above.

The initial binary population is summarized in Figure \ref{fig:init_binary_pop}, using an $N=10^6$ stellar population sampled with \texttt{COSMIC} to reduce noise in the histograms and aid in visualization. The binary fraction is $f_{\rm bin}\sim0.41$ for solar-type primaries, and increases to $1$ at $40~M_{\odot}$. Below $<0.8 M_{\odot}$ (where \citealt{moe2017} do not prescribe binary demographics), the \texttt{COSMIC} binary fraction decreases linearly with $\log M_1$ to 0 at $M_1=0.08 M_{\odot}$. 
We show the period, eccentricity, and mass ratio distributions separately for systems with primaries in several mass ranges, motivated by the breaks in the piecewise demographic prescriptions. For solar-type and lower-mass primaries, binaries follow lognormal period distributions that peak at $\sim10^3$~yr, while at higher masses, binaries have flatter distributions skewed toward shorter periods (see also Figure 37 in \citealt{moe2017}). 
The eccentricity distributions are near-thermal, with peaks of circularized systems at short periods. The mass ratio distributions follow double power laws, with excesses of equal-mass systems. 

Our main simulation grid does not vary the initial binary population, so we probe the sensitivity of our results to the initial binary fraction with two additional simulations. In the first, we maintain the shapes of the binary demographic distributions from \citet{moe2017}, but double the overall binary fraction. Because the binary fraction increases with primary mass, this yields a cluster with an unrealistically top-heavy IMF, and a number of OB stars similar to the hi-OB population from the main grid. To remove this effect in the second additional simulation, we double the binary fraction only for primaries $<1.2~M_{\odot}$, leaving the demographic distributions otherwise unchanged. This draw yields a fraction of OB stars similar to the lo-OB population, and a nearly doubled binary fraction of $f_{\rm bin}\sim0.4$ compared to the main grid. 
The initial conditions of the clusters with altered binary populations are listed in the bottom rows of Table \ref{tab:simulation_grid}; they are placed on GD-1 orbits and have $R_{\rm vir,0}=1.5~\rm pc$.

After generating the stellar population, we temporarily replace each binary with the combined mass of the two companion stars for placement in the multi-mass King profile. We then restore two separate stars in an orbit of the correct eccentricity and period about a center of mass with the appropriate position and velocity in the King profile. Finally, we reorient each binary's angular momentum along a vector drawn from a $\cos{i}$ isotropic distribution. Our routine to initialize binary orbits and place them into the cluster distribution function is based on the \texttt{kira.py} script from the examples provided with the AMUSE code base \citep{amuse2009,amuse2013}. 

\begin{figure*}
    \centering
    \includegraphics[width=\linewidth]{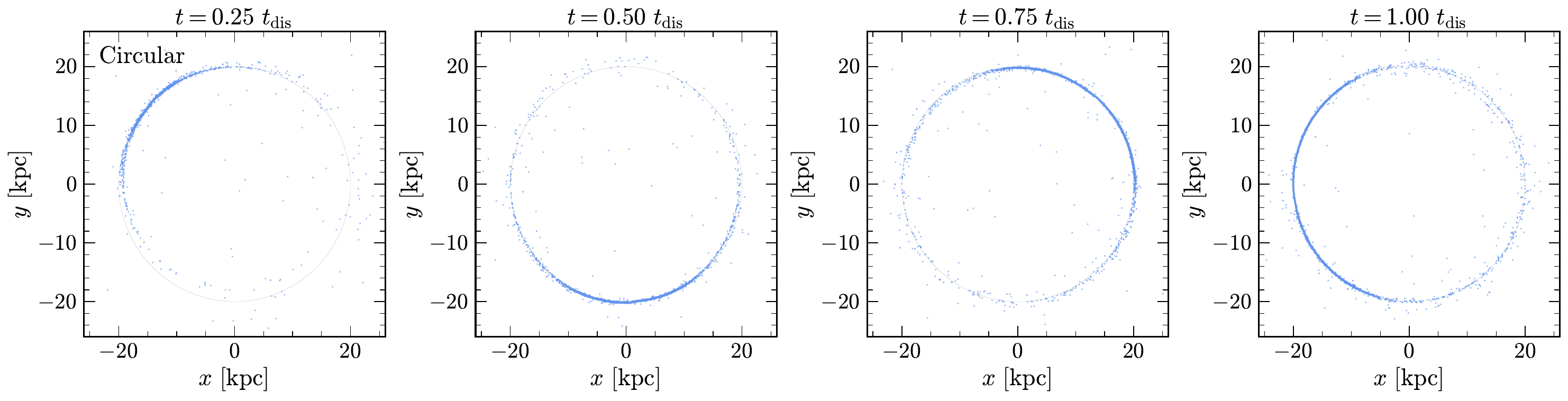}
    \includegraphics[width=\linewidth]{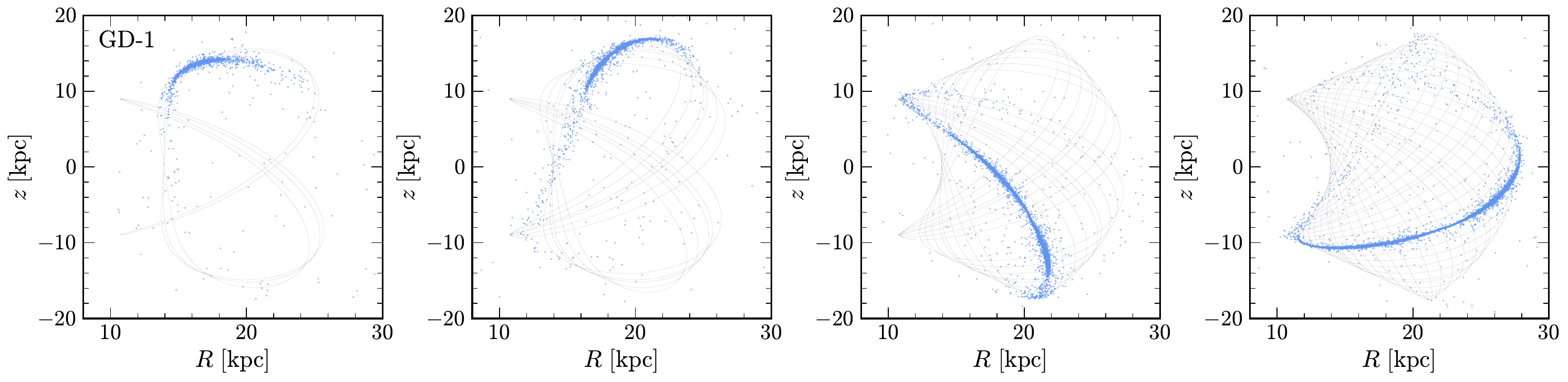}
    \includegraphics[width=\linewidth]{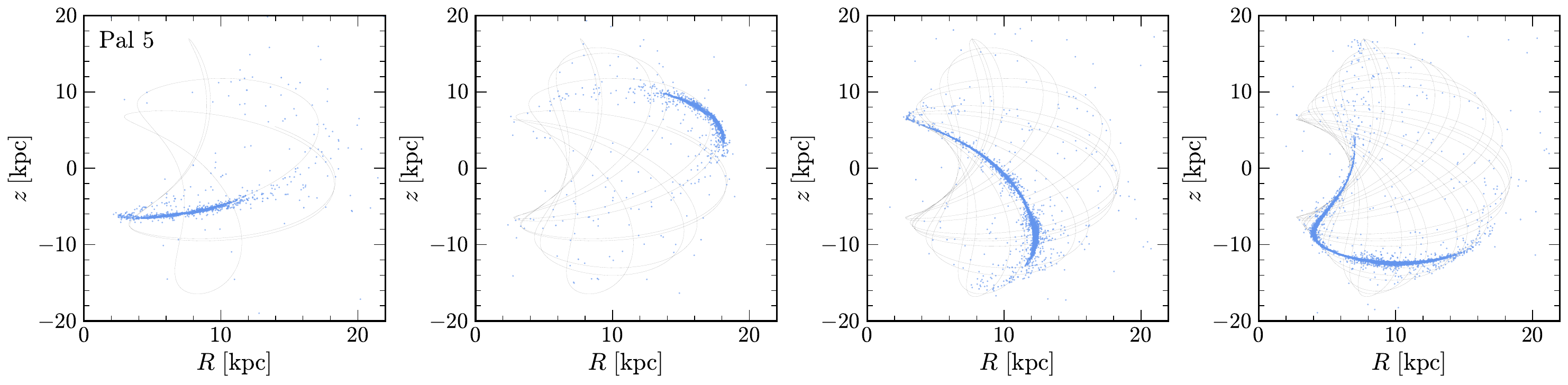}
    \caption{Snapshots over the course of the evolution of each lo-OB cluster with $R_{\rm vir,0}=0.75$~pc. The first, second, and third rows of panels display clusters on circular, GD-1, and Pal 5-like orbits, respectively. Successive columns display the stream at later stages of the progenitor cluster's dissolution. The circular orbit is shown in the $x\text{-}y$ plane, while the eccentric orbits are shown in the meridional plane ($z$ vs $R=\sqrt{x^2 + y^2}$) to best showcase the orbital structure.}
    \label{fig:stream_dissolution}
\end{figure*}
We compute the initial half-mass relaxation time $t_{rh,0}$ for each cluster using Equation \eqref{eq:t_relax} and provide it in Table \ref{tab:simulation_grid}. We set the crossing time equal to the local dynamical timescale at the half mass radius, taking the density to be the average density of a spherical shell around the cluster with thickness $r_h\pm0.5$ pc (this initial dynamical time $t_{\rm dyn,0}$ is also listed in Table \ref{tab:simulation_grid}). We also approximate $\ln{\Lambda}\approx\ln{N}$. 
Finally, we compute and provide in Table \ref{tab:simulation_grid} the intrinsic initial 1D velocity dispersion, given by
\begin{equation}
    \label{eq:dispersion}
    \sigma_{\rm 1D} = \sqrt{\frac{\langle v^2 \rangle-\langle v\rangle^2}{3}}.
\end{equation}
We use the systemic velocities of the binaries, rather than each star's individual velocity, when computing velocity dispersions.

\begin{figure*}
    \centering
    \includegraphics[width=\linewidth]{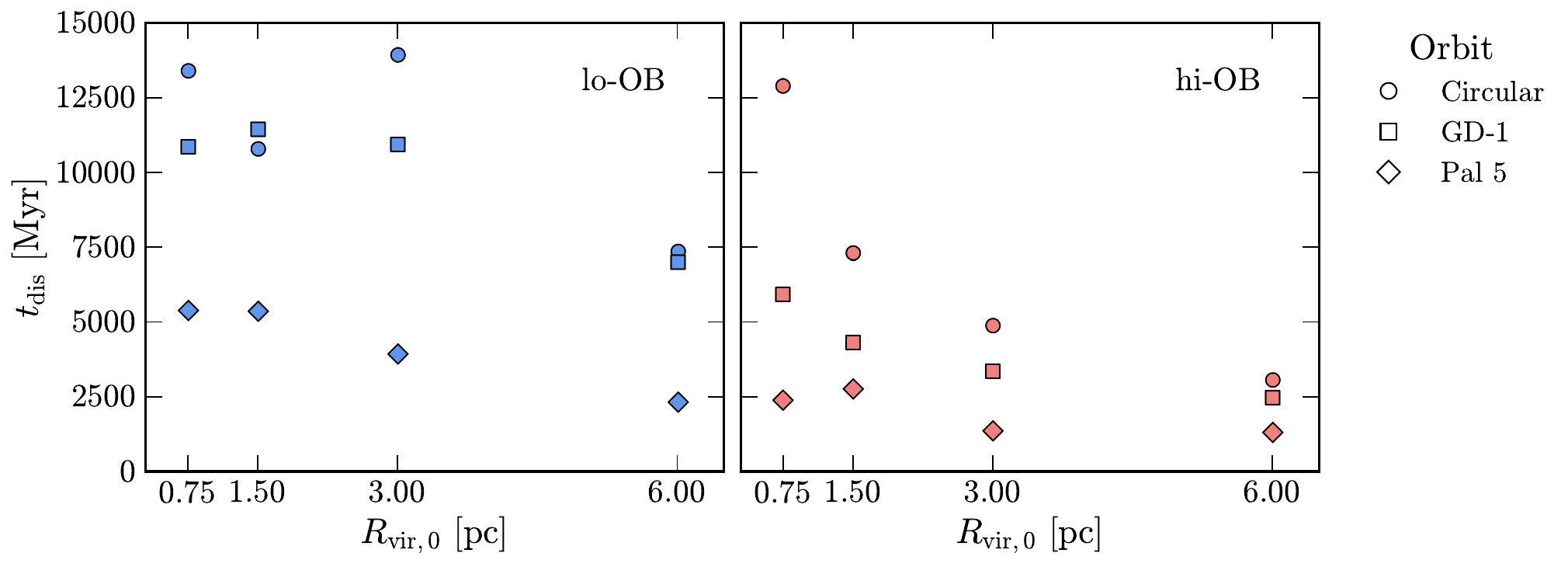}
    \caption{Dissolution time of each simulated cluster listed in Table \ref{tab:simulation_grid} as a function of initial virial radius for lo-OB (left panel) and hi-OB (right panel) simulations. Marker shapes correspond to the progenitor orbit.}
    \label{fig:tdis_summary}
\end{figure*}

\begin{figure*}
    \centering
    \includegraphics[width=\textwidth]{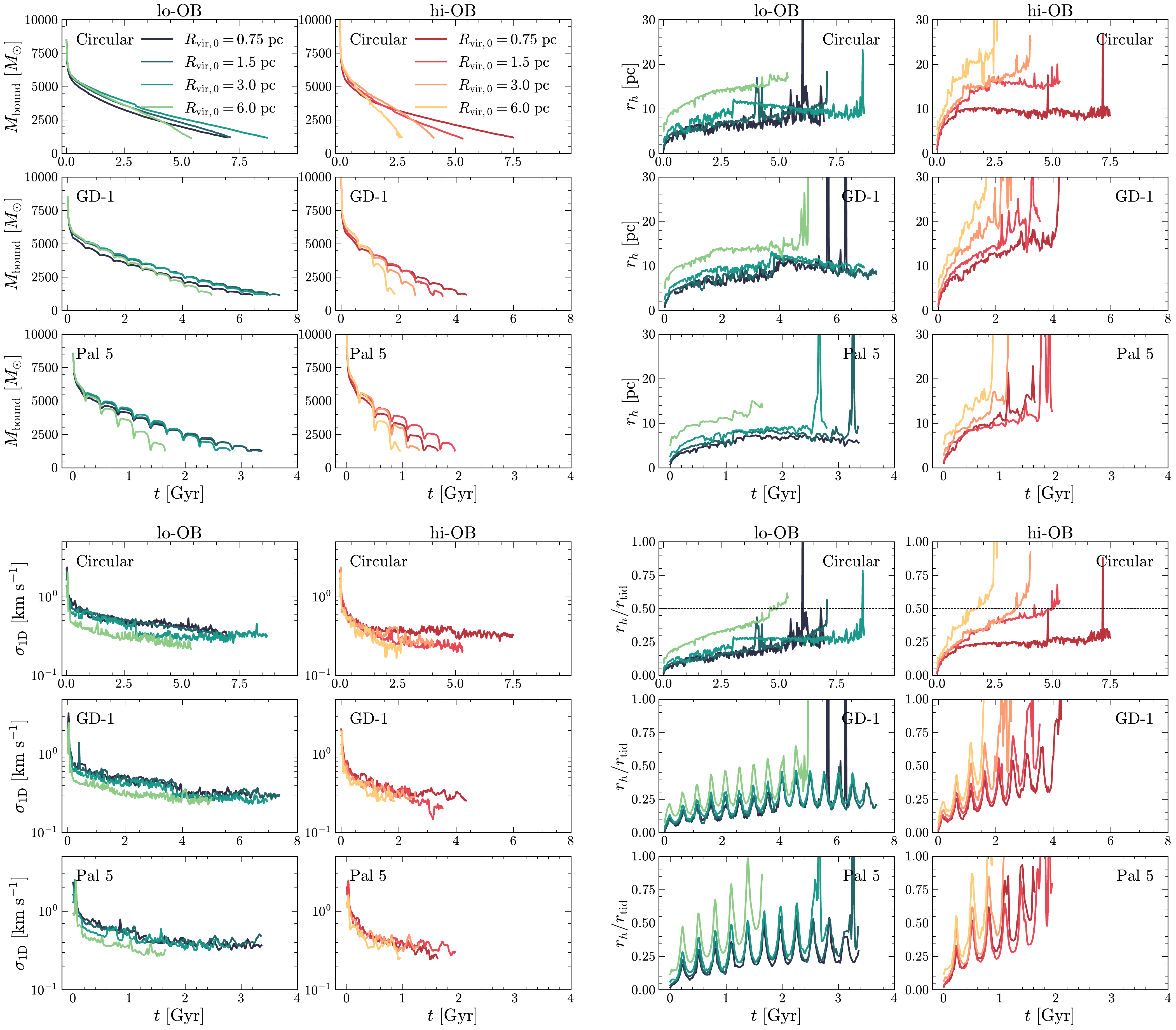}
    \caption{Evolution of each simulated cluster in bound mass (top left quadrant), half mass radius (top right quadrant), velocity dispersion (bottom left quadrant), and tidal filling factor (bottom right quadrant). Within each quadrant, the results for the low (high) mass simulations are shown in the left (right) column, and the clusters on circular, GD-1, and Pal 5 orbits are shown in the top, center, and bottom rows, respectively. Data from initially denser clusters are shown in darker shades.}
    \label{fig:structural_evolution}
\end{figure*}

\subsection{N-body Implementation}\label{sec:methods_2_impl}
We simulate clusters with the collisional $N$-body simulation code \textsc{petar} \citep{Wang2020-petar}. \textsc{petar} employs a fourth-order Hermite integrator for short-range forces, and a \citet{barneshut1986} particle tree for long range forces through the simulation library \textsc{fdps} \citep{IwasawaFDPS2016, IwasawaFDPS2020}. To evolve clusters with arbitrarily large binary fractions in a reasonable wall-clock time, \textsc{petar} applies the slow-down algorithmic regularization technique (\textsc{sdar}) described in \citet{wang2020-sdar} for binary orbits and hyperbolic encounters. 

\textsc{petar} accounts for stellar evolution using the prescriptions of the SSE/BSE stellar evolution packages \citep{SSE_2000,BSE_2002}. We configure \textsc{petar} with the updated SSE/BSE versions described in \citealt{BSE2020}, and use a stellar metallicity $\log{Z/Z_{\odot}}=-2$. 
For the Galactic potential, we use the axisymmetric default potential \texttt{MWPotential2014} from \texttt{galpy} \citep{bovy2015-galpy}, which includes three components: a spherical power law bulge potential, an axisymmetric \citet{miyamoto1975-potential} disk potential, and a \citet{NFW1997} dark matter halo potential. Specific parameters for each component are fit to match Milky Way observations in \citet{bovy2015-galpy}. 

Each simulation begins using \textsc{petar}'s automatically-determined tree time step and changeover radius. These parameters prescribe the regimes in which \textsc{petar} integrates particle motions using direct summation of forces versus using the Barnes-Hut particle tree (we direct the reader to the software documentation\footnote{\href{https://github.com/lwang-astro/PeTar}{https://github.com/lwang-astro/PeTar}} for further details). 
\textsc{petar}'s performance is sensitive to the tree time steps \citep{wang2026}. For the most diffuse clusters in our simulation grid, the wall-clock time to simulate the cluster until dissolution using the automatically-determined values at the initial condition is $\lesssim$1 week. However, for the densest clusters, the simulation runtime using the automatically determined values at the initial condition is prohibitively slow (multiple weeks of wall-clock time are required to simulate 1 Gyr of cluster evolution). With a lower density and velocity dispersion, the tree time steps can be adjusted to larger values to speed up the simulations, so we use \textsc{petar}'s \texttt{find.dt} functionality to pause the simulations after a few weeks of wall clock time (or several Gyr of simulation time), and resume them using a custom, optimized tree timestep following significant changes to the cluster structure in its early evolution (see Section \ref{sec:results_1_structure}). We follow this procedure for all simulations with $R_{\rm{vir},0}=0.75$~pc, which we simulated until cluster dissolution in on average 10 weeks of wall clock time. 

Each simulation is run until the cluster dissolution time $t_{\rm{dis}}$, which we define as the last time when $>100$ stars remain within the tidal radius. 
We provide the dissolution time for each cluster in Table \ref{tab:simulation_grid}, as well as the time of their last pericenter passage before dissolution, determined using the last minimum in the radial Galactocentric position of the cluster before $t_{\rm{dis}}$. Streams are longest and most symmetric at pericenter (see e.g., \citealt{hozumiDevelopmentMultipleTidal2015}), so we choose this time to display them in Section \ref{sec:results_3_streams}, mainly for aesthetic purposes. 

We parallelize each simulation over 8 CPU cores through \textsc{petar}'s use of \textsc{OpenMP}. The parallelization scheme is dynamic, so that the assignment of computations to different cores is nondeterministic. The accumulation of floating-point round off errors is therefore stochastic, and ultimately can result in 
significantly different simulation outcomes, even using identical initial conditions (see, e.g., \citealt{hoffmannStochasticityPredictabilityTerrestrial2017, kellerChaosVarianceGalaxy2018}). 
To probe the issue empirically, we ran 10 realizations of the lo-OB and hi-OB clusters on GD-1 orbits with $R_{\rm vir,0}=0.75~\rm pc$ and $R_{\rm vir,0}=6.0~\rm pc$. In the final two columns of Table \ref{tab:simulation_grid}, we report the $16^{\rm th}$--$84^{\rm th}$ percentile range in dissolution times to capture the variance in outcomes. The dissolution times for initially identical clusters vary as much as 10--20\%.

\subsection{Data processing}\label{sec:methods_3_analysis}
At the end of each simulation, we use \textsc{petar}'s built-in data processing routines to identify single and binary stars at each output. We ignore any higher-order multistellar systems that may form during the simulations in this work. The data processing routines compute the location of the simulation's density peak at each time step, or the ``core," which we use to define the cluster reference frame (e.g., the cluster location when determining the time of the last pericenter passage). Finally, \textsc{petar} automatically determines the tidal radius (using $f=3$ in Equation \ref{eq:rtid}). 

We use the half mass radius $r_h$ to characterize the size of the bound cluster over time. In order to estimate the cluster size independently of the tidal radius (so as to compare the two appropriately in Section \ref{sec:results}), we approximate $r_h$ by finding the radius enclosing half the mass of stars that are energetically bound to the cluster. We neglect the external potential and the rotating reference frame of the cluster in this calculation, but this assumption breaks down when few stars remain within the cluster. In Section \ref{sec:results} we display $r_h$ (and $r_h/r_{\rm tid}$) only while $>2500$ stars remain within the tidal radius, choosing this criterion empirically based on the behavior of the resulting $r_h$ calculations, which become increasingly numerically unstable as stars escape the cluster. 

In Section \ref{sec:results}, we examine the hard/soft boundary for individual binaries in the simulated clusters. To compute $a_{\rm h/s}$ with Equation \eqref{eq:ahs} for a binary a distance $r$ from the cluster core, we find the locally averaged stellar mass $\langle m \rangle$ and local 3D (i.e., without the factor of $1/\sqrt{3}$ in Equation \eqref{eq:dispersion}) velocity dispersion $\sigma$ in a spherical shell with radial boundaries $r\pm 1$ pc (unless the binary is $<1$ pc from the cluster center, in which case we use the averaged mass and dispersion within the sphere of radius $r+1$ pc).

To display the streams at the last pericenter passage before dissolution, we transform from Galactocentric to ``stream frame" coordinates.\footnote{The script we use for transformation to stream coordinates is available at \href{https://github.com/jnibauer/streamframe}{https://github.com/jnibauer/streamframe}.} 
The Galactocentric coordinates are first transformed in position and velocity to align the $z$ axis with the angular momentum vector of the progenitor. The $x$ axis points from the Galactic center to the progenitor position, and the $y$ axis is given by the cross product $\hat{z}\times \hat{x}$. The rotated cartesian coordinates are then transformed to stream frame longitude $\phi_1$ and latitude $\phi_2$ by 
$\tan\phi_1=y/x$, 
$\phi_2=\arcsin(z/r)$, where $r=\sqrt{x^2+y^2+z^2}$. We also obtain the radial velocity $v_r$ relative to the galactic center. 
To compute the progenitor angular momentum, we use the position and velocity of the cluster core returned in \textsc{petar}'s post-processing routines.

We discard stars with radial coordinates larger than $1.5~r_{\rm{apo}}$ given the apocenter of the cluster orbit to roughly remove stars that are not part of the stream. We then trim the remaining stars using a $3\sigma$ clipping in $\phi_1,\ \phi_2$, and $r$. Finally, we fit and subtract a fifth-order polynomial to the stream in $\phi_2$ versus $\phi_1$ in order to straighten it in longitude and latitude. This is the lowest-order fit that we find to consistently yield well straightened streams across the simulation grid.  
\begin{figure}
    \centering
    \includegraphics[width=\linewidth]{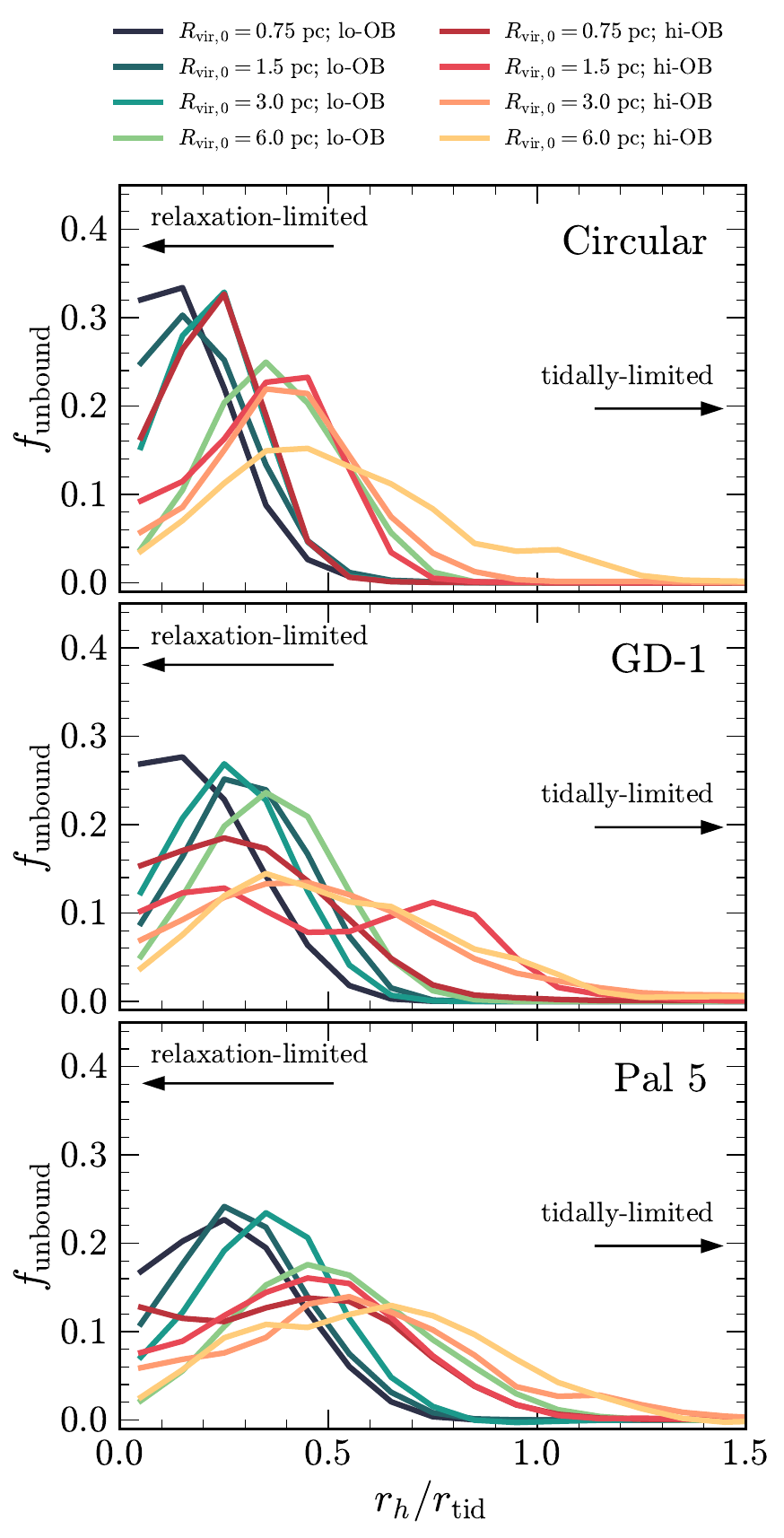}
    \caption{Fraction of stars unbound from each cluster as a function of the cluster tidal filling factor at the time of their unbinding. The top, center, and bottom panels show curves for clusters on circular, GD-1, and Pal 5 orbits, respectively. The shape and peak of each distribution reflects the underlying ejection mechanism responsible for the cluster's dissolution: the dissolution of higher density and lo-OB clusters is more relaxation-limited, while the dissolution of lower-density and hi-OB clusters is more tidally-limited.}
    \label{fig:rh/rtid_hist}
\end{figure}

\begin{figure*}
    \centering
    \includegraphics[width=\linewidth]{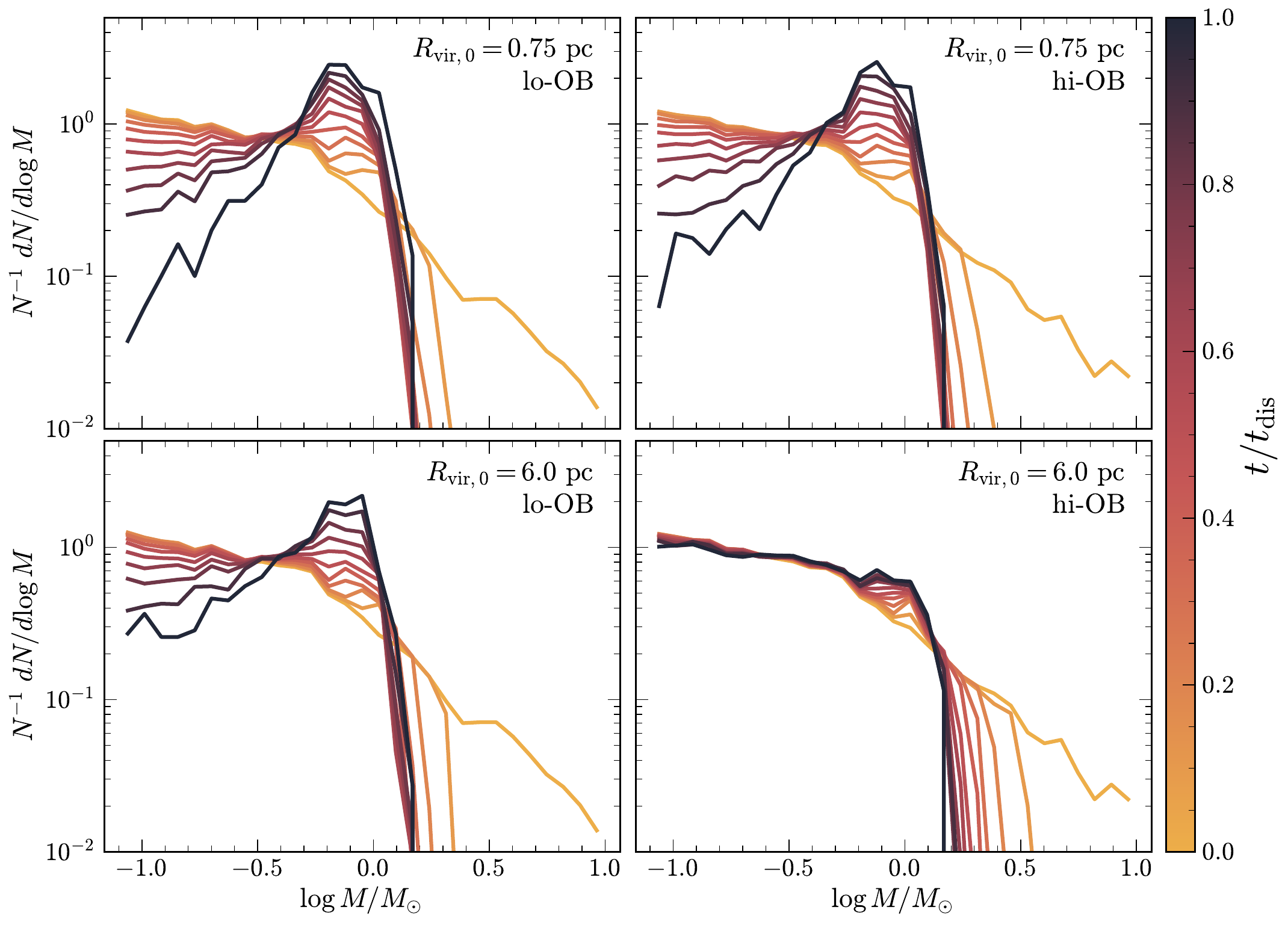}
    \caption{Mass functions of stars within the cluster tidal radius over time, from the initial condition to the dissolution time in steps of $t_{\rm{dis}}/10$. The MF includes both of the companion masses in each binary as well as remnant objects. The simulations shown are the initially most dense and least dense lo-OB and hi-OB clusters on GD-1 orbits. MFs at later time steps are shown in darker shades. To reduce shot noise, we stack the results from the ten realizations of each simulation described in Section \ref{sec:methods_2_impl} at consistent $t/t_{\rm dis}$.}
    \label{fig:bound_MFs}
\end{figure*}
\begin{figure*}
    \centering
    \includegraphics[width=\textwidth]{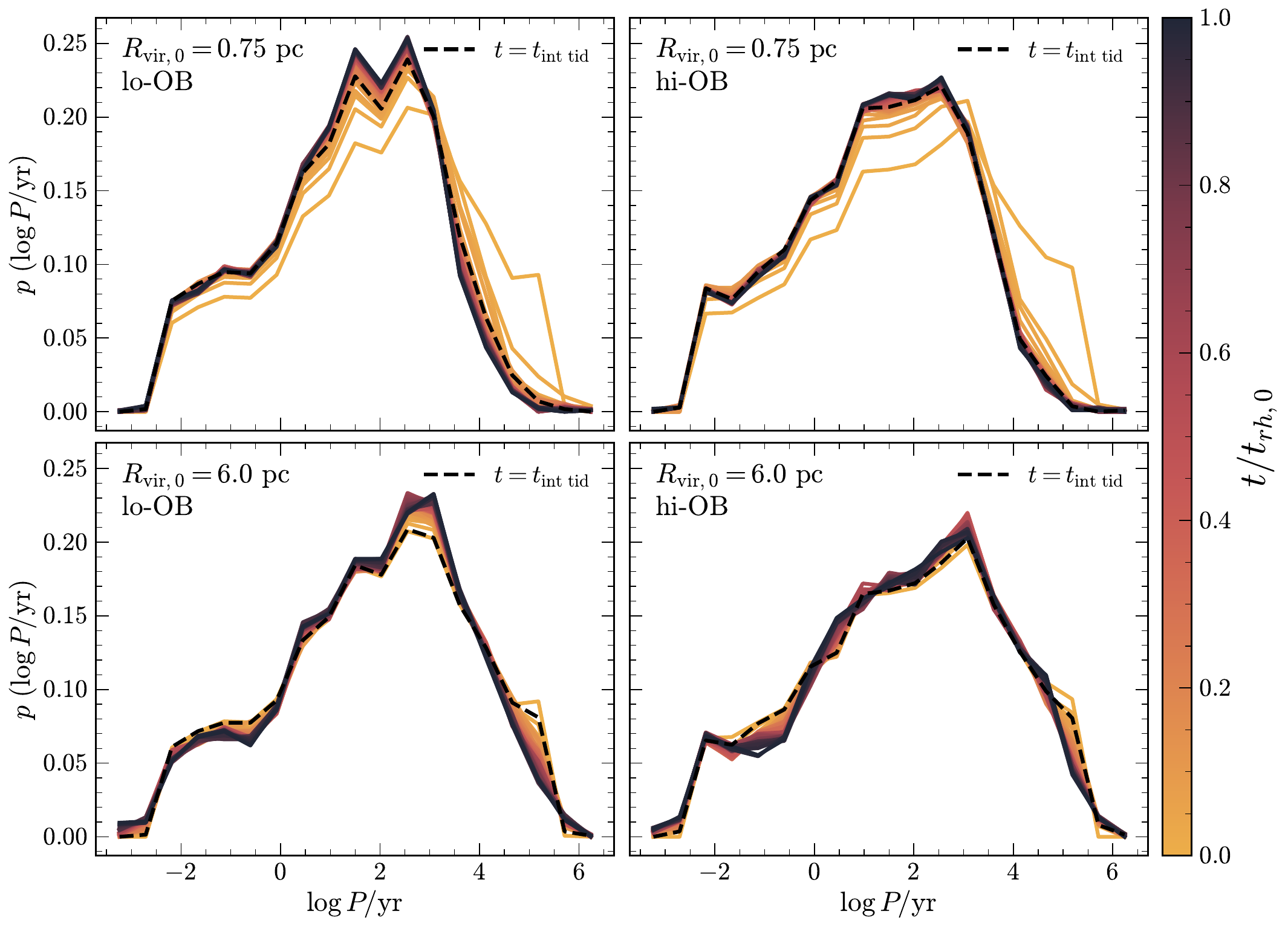}
    \caption{Distribution of orbital periods for binaries within the tidal radius over time from $t=0$ to the initial half mass relaxation time $t_{rh,0}$ in steps of $t_{rh,0}/50$, expressed as probability densities. Distributions at later time steps are shown in darker shades, and dashed black histograms show the period distribution after cluster expansion following the deaths of $\gtrsim25~M_{\odot}$ stars. The simulations shown are the same as those in Figure \ref{fig:bound_MFs}.}
    \label{fig:pDists}
\end{figure*}
\begin{figure*}
    \centering
    \includegraphics[width=\linewidth]{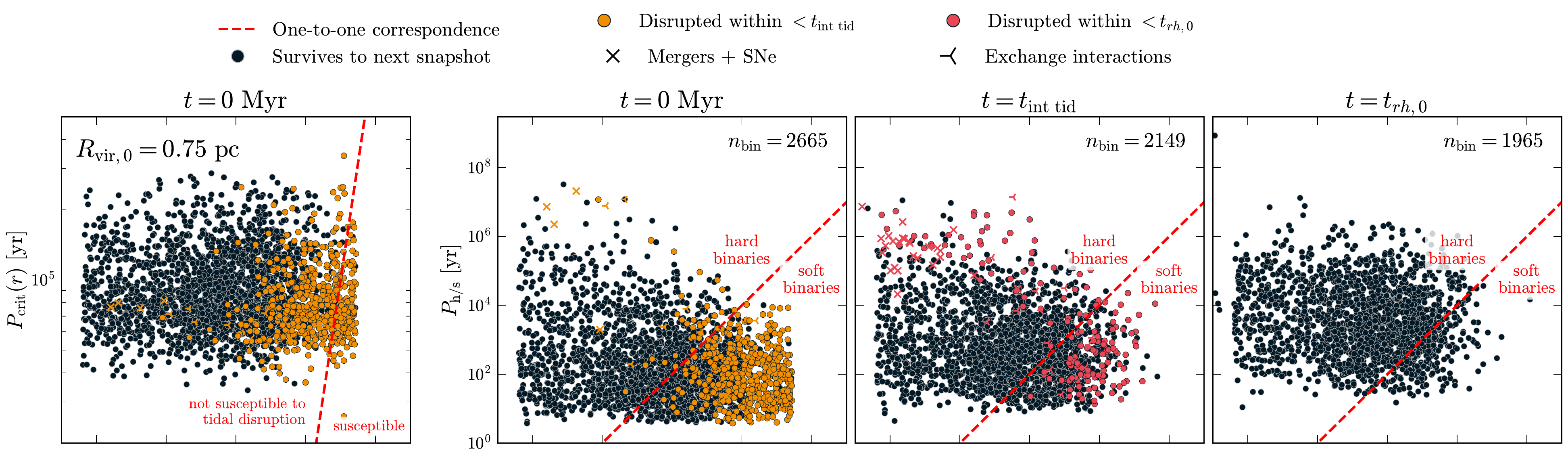}
    \includegraphics[width=\linewidth]{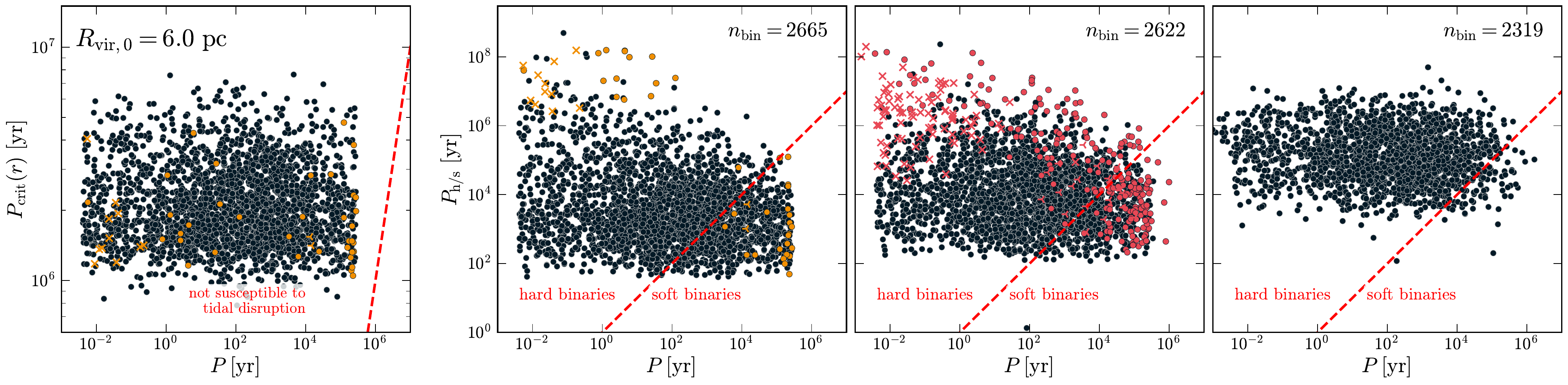}
    \caption{An exploration of the depletion of wide binaries in the most dense (top panels) and least dense (bottom panels) lo-OB clusters on GD-1-like orbits. Orange markers highlight binaries that disappear before $t=t_{\rm int\ tid}$ (the timescale for cluster expansion following the deaths of $\geq25~M_{\odot}$ stars). Pink markers highlight binaries that disappear between $t_{\rm int\ tid}$ and the first half mass relaxation time. We distinguish binaries that disappear due to true disruptions, mergers, and exchange interactions with distinct markers. \textbf{Leftmost panels}: the critical period above which a binary can be disrupted by tides internal to the cluster versus the actual orbital periods at $t=0$. The red dashed line shows the one-to-one correspondence so that systems below and to the right of it can be disrupted by tides. \textbf{Right three sets of panels}: the hard/soft boundary versus the true orbital periods at three time steps of the simulation---(from left to right) at $t=0$, after $t=t_{\rm int\ tid}$, and after the first half mass relaxation time. The red dashed line shows the one-to-one correspondence so that systems below and to the right of it are soft binaries and can be disrupted by gravitational interactions with other stars in the cluster. In the top right corners, we show the number of binaries present in the simulation at each timestep. By $t_{rh,0}$, soft binaries are strongly depleted while hard binaries remain resilient to tides and two-body encounters.}
    \label{fig:roche_hs}
\end{figure*}
\begin{figure}
    \centering
    \includegraphics[width=0.888\linewidth]{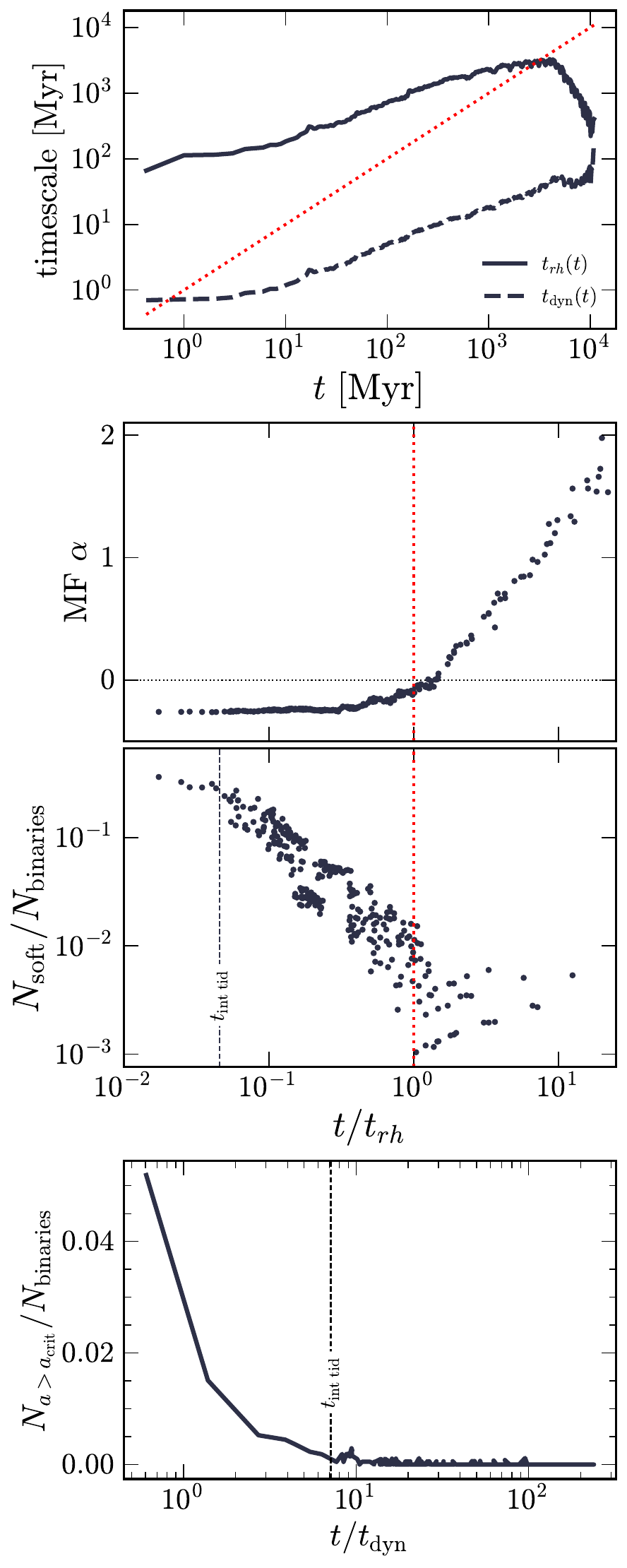}
    \caption{Structural and soft binary evolution of the densest lo-OB cluster on a GD-1 orbit, showing the evolving $t_{rh}(t)$ and $t_{\rm dyn}(t)$ (top panel), the low mass MF slope and soft binary fraction over time in units of $t_{rh}(t)$ (central two panels), and the fraction of binaries susceptible to disruption by internal tides over time in units of $t_{\rm dyn}(t)$ (bottom panel). Thick dotted red lines mark where the present time $t$ is equal to the half mass relaxation time $t_{rh}(t)$, and thin dashed black lines mark the timescale for cluster expansion following the deaths of $\gtrsim25~M_{\odot}$ stars.}    
    \label{fig:depleting_longperiods_late}
\end{figure}

\section{Results}\label{sec:results}
We now turn to the simulation outcomes. We examine the structural evolution of each cluster in Section \ref{sec:results_1_structure} and the evolution of the stellar and binary populations within the clusters in Section \ref{sec:results_2_stellarpops}. We present the resulting streams in Section \ref{sec:results_3_streams} and perform a mock RV survey to determine a realistic binary detection rate in Section \ref{sec:results4_mocksurvey}. 
Figure \ref{fig:stream_dissolution} shows the formation of streams for the densest lo-OB clusters for each simulated orbit type. 

\subsection{Cluster structural evolution}\label{sec:results_1_structure}
We turn to the evolution of the cluster stellar population in Figure \ref{fig:tdis_summary}, which shows the dissolution time for each simulation in our grid given its initial virial radius and cluster orbit. 
The dissolution times decrease with increasing initial virial radius (and decreasing initial cluster density). 
Hi-OB clusters are unbound earlier than lo-OB clusters due to their enhanced mass loss and expansion from massive stellar evolution. 
Clusters on Pal 5 orbits experience the strongest tidal fields due to Pal 5's small pericenter, and therefore are the earliest to dissolve, typically followed by clusters on GD-1 orbits, and finally clusters on circular orbits. 
We speculate that the simulations that behave differently (for example, the unusually short dissolution time of lo-OB cluster with $R_{\rm vir,0}=1.5~\rm pc$ on a circular orbit) are due to stochastic effects due to the chaotic nature of the system and from \textsc{petar}'s dynamic parallelization scheme (see Section \ref{sec:methods_2_impl}). Short relaxation timescales in the densest lo-OB clusters may also act to decrease the dissolution time (e.g., \citealt{binneyTremaine,weatherford2024}).

Figure \ref{fig:structural_evolution} shows the structural evolution of each cluster until its dissolution. We show the mass within the tidal radius $M_{\rm{bound}}$, the half mass radius $r_h$, the 1D velocity dispersion $\sigma_{\rm 1D}$ of stars within the tidal radius, and the ratio of the half-mass and tidal radii $r_h/r_{\rm{tid}}$. For all quantities except $M_{\rm bound}$, we apply a Gaussian smoothing filter with $\sigma=10$~Myr and truncating at $\pm4\sigma$ to remove noisy behavior in our calculation of $\sigma_{\rm 1D}$ and $r_h$. 
All clusters show rapid mass loss from stellar evolution at early times, and a corresponding increase in size and decrease in velocity dispersion. It is common for clusters to be tidally-filling, with $r_h/r_{\rm tid}\gtrsim0.5$, toward the end of their lifetimes, and for clusters on eccentric orbits to periodically reach this threshold with the oscillation of their tidal radii over the course of their orbits.

We further investigate the loss of stars from each cluster as a function of its tidal filling factor $r_h/r_{\rm tid}$ in Figure \ref{fig:rh/rtid_hist}. We find the change in the count of stars bound to the cluster at each timestep, along with the corresponding tidal filling factor. We then sum the changes to the number of bound stars in bins of the tidal filling factor to find the net $\Delta N_{\rm bound}$ as a function of $r_h/r_{\rm tid}$. We express values in each bin as the fraction of stars unbound from the cluster in each $r_h/r_{\rm tid}$ bin, $f_{\rm unbound}=\Delta N_{\rm bound,\ bin}/\Delta N_{\rm bound,\ total}$. 
To suppress noise, we include only times where the number of bound stars exceeds 2500 and the tidal radius has changed from the previous time step by $<1~\rm pc$. The latter step removes numerical artifacts in the tidal radius from \textsc{petar}'s data processing routine. We apply another Gaussian smoothing filter with $\sigma=0.1$ (equal to the bin width of the histograms).

For a given OB stellar population and orbit shape, the peak of stars lost at small $r_h/r_{\rm tid}$ in the relaxation-limited regime is highest for the initially densest clusters, with more diffuse clusters losing more stars in the tidally-limited regime. For a given initial cluster density and orbit, dissolution is more relaxation-limited for the lo-OB clusters and more tidally-limited for the hi-OB clusters. Finally, for a given density and OB population, the dissolution of clusters on circular orbits is the most relaxation-limited, and the dissolution of clusters on Pal 5 orbits is most tidally-limited. 
Some clusters, most notably the hi-OB cluster with $R_{\rm vir, 0}=1.5~\rm pc$ on a GD-1 orbit, exhibit bimodal histograms in Figure \ref{fig:rh/rtid_hist}, with peaks of dissolution in more relaxation-limited and more tidally-limited regimes. 

\subsection{Evolution of the cluster stellar population}\label{sec:results_2_stellarpops}
Figure \ref{fig:bound_MFs} shows the evolution of the stellar mass function within the tidal radius for four illustrative simulations in our grid: the lo-OB and hi-OB cases of the most and least dense clusters on GD-1 orbits. 
We stack mass functions from time snapshots at consistent $t/t_{\rm dis}$ across the ten realizations of each simulation described in Section \ref{sec:methods_2_impl} to minimize noisy behavior in mass ranges where single simulations contain few stars. The masses of stars in binaries are included separately, rather than as the binary's combined mass. Remnant objects are also included in the data, and we note that the clusters shown have remnant fractions of $\sim10$--$50\%$ by number near their dissolution times (where the remnants are dominated by white dwarfs and the diffuse clusters have lower remnant fractions).
All cluster MFs are quickly truncated at high masses due to mass loss from stellar evolution. The slope of the MF at low masses increases throughout the simulation as the cluster mass segregates and low mass stars are preferentially lost through evaporation. The degree of mass segregation is most dependent on the initial cluster density. Dense clusters with short relaxation times mass segregate quickly and have longer lifetimes, so that their low mass MF slopes show larger increases by $t=t_{\rm dis}$ compared to diffuse clusters with long relaxation times. 

Figure \ref{fig:pDists} shows the evolution of the binary period distribution within the tidal radii of the same four clusters as are shown in Figure \ref{fig:bound_MFs}. The distributions are shown during the first initial relaxation time, in steps of $t_{rh,0}/50$. 
We include a curve showing the period distribution after $t_{\rm int\ tid}=t_{\rm SE}(25~M_{\odot}) + t_{\rm dyn}$, capturing the predicted timescale from Equation \eqref{eq:t_internaltides} for the expansion of the cluster after impulsive mass loss from massive stellar evolution. The timescale $t_{\rm int\ tid}$ matches the observed timescale for rapid disruption of the widest binaries in the cluster, after which the changes to the period distributions are more gradual. 
The densest clusters significantly deplete wide binaries with $\log{P/\rm yr}>3$, whereas for the most diffuse clusters, the boundary is closer to $\log{P/\rm{yr}}>5$. 
There is little distinction in the evolving behavior of the period distribution between lo-OB and hi-OB clusters, indicating that wide binary depletion is set primarily by the initial cluster density, rather than the exact expansion factor after mass loss from stellar evolution.

We further investigate the disruption of wide binaries in the early evolution of the most and least dense lo-OB clusters on GD-1 orbits as examples in Figure \ref{fig:roche_hs}. We compute the critical orbital period at which each binary can be disrupted by differential tidal forces interior to the cluster itself (Equation \ref{eq:roche_limit}) at $t=0$ and show the distribution in $P_{\rm crit}$ (converted from $a_{\rm crit}$ using Kepler's 3rd law) against the actual orbital periods at $t=0$ in the leftmost panels. We highlight systems that are disrupted at times $<t_{\rm int\ tid}$. For the denser cluster, many of these systems have $P>P_{\rm crit}$.

The right set of panels in Figure \ref{fig:roche_hs} show the orbital period at the hard/soft boundary against the true periods. At $t=0$, we highlight again the systems disrupted within $<t_{\rm int\ tid}$ of the simulation. For both clusters, nearly all of these systems are soft binaries. 
Next, we show the same distribution at $t=t_{\rm int\ tid}$ and we highlight the systems that are disrupted within the first initial half-mass relaxation time. The majority of these disrupted systems are soft binaries. 
We distinguish ``disrupted" systems (in which the two binary members appear at the next timestep as single stars), ``mergers" (in which one or both of the binary members is completely absent from the simulation at the next time step), and ``exchange interactions" (in which one or both binary members appear in different binary systems at the next time step). The disappearing hard binaries are often merger products and exchange interactions rather than true disruptions. 
Finally, in the rightmost panel of Figure \ref{fig:roche_hs}, we show all surviving binaries after the initial half mass relaxation time has passed, demonstrating that relatively few soft binaries remain in the cluster.

A detailed summary of the structural and soft binary evolution of the densest lo-OB cluster on a GD-1 orbit is provided in Figure \ref{fig:depleting_longperiods_late}. The top panel shows the evolution of the cluster half mass relaxation and dynamical timescales throughout the simulation. The relaxation timescale increases at early times $t<t_{rh}(t)$ due to the expansion and decreasing density of the cluster, and decreases at late times $t>t_{rh}(t)$ due to the decreasing number of bound stars. 

In the central panels, we show two quantities as a function of time in units of the present relaxation time $t_{rh}(t)$. First, we show the power law index of the cluster MF between $-0.5<\log{M/M_{\odot}}<-0.3$. We use a finite difference approximation to the MF slope across the endpoints of the mass window after binning the mass data in equal-width bins with centers at $\log{M/M_{\odot}}= -0.5,-0.4,-0.3$. The onset of mass segregation is, as expected, governed by the relaxation timescale, and occurs at $t\approx t_{rh}(t)$ for both clusters. 
Second, we show the fraction of soft binaries, with $a>a_{\rm h/s}$. Although a minority of soft binaries remain in the clusters at $t=t_{rh,0}$ in the right panels of Figure \ref{fig:roche_hs}, they are depleted to $\ll 1\%$ by the time $t=t_{rh}(t)$.

Finally, in the bottom panel, we show the fraction of binaries susceptible to disruption by internal tides ($a>a_{\rm crit}$), as a function of time in units of $t_{\rm dyn}(t)$. The depletion of tidally-disruptable binaries corresponds to $t=t_{\rm int\ tid}$. We also mark $t_{\rm int\ tid}$ in the panel above, showing that the slight change in slope in the decreasing fraction of all soft binaries coincides with the depletion of those susceptible to disruption by internal tides.

\subsection{Binaries in the resulting streams}\label{sec:results_3_streams}
\begin{figure*}
    \centering
    \includegraphics[width=\textwidth]{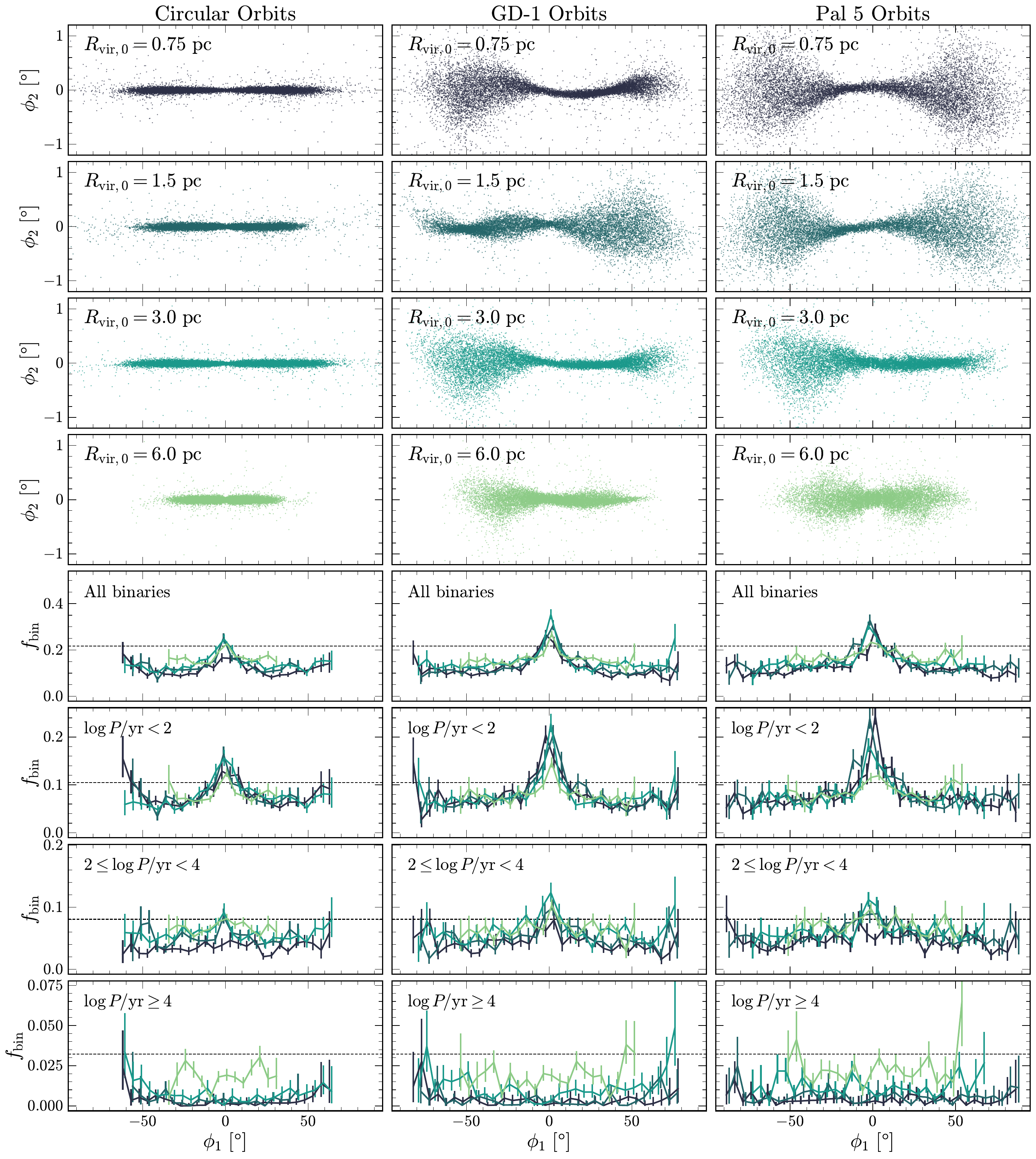}
    \caption{Summary of stream morphology and binary fractions at the last pericenter passage before dissolution for the lo-OB simulations. The colors corresponding to each simulation are the same as those in Figure \ref{fig:structural_evolution}, so that initially denser clusters are shown in darker shades. The streams on circular, GD-1, and Pal 5-like orbits are shown in the left, center, and right column of panels, respectively. \textbf{Top 12 panels}: the resulting streams from each lo-OB cluster at their last pericenter passage before dissolution in stream coordinates $\phi_2$ versus $\phi_1$. \textbf{Bottom 12 panels}: at the same time, the total binary fraction (topmost row), or the binary fraction for periods $P<100$ yr, $100\leq P<10^4$ yr, or $P\geq10^4$ yr (bottom three rows) in $5\degree$ bins along $\phi_1$. The binary fraction in a period range is defined as $N_{\rm{binaries\ in\ period\ range}}/N_{\rm{systems}}$. We show only bins that contain $\geq 50$ systems in order to mitigate noise. Horizontal dashed lines show the initial value of the binary fractions.  
    }
    \label{fig:fbin_lm}
\end{figure*}
\begin{figure*}
    \centering
    \includegraphics[width=\textwidth]{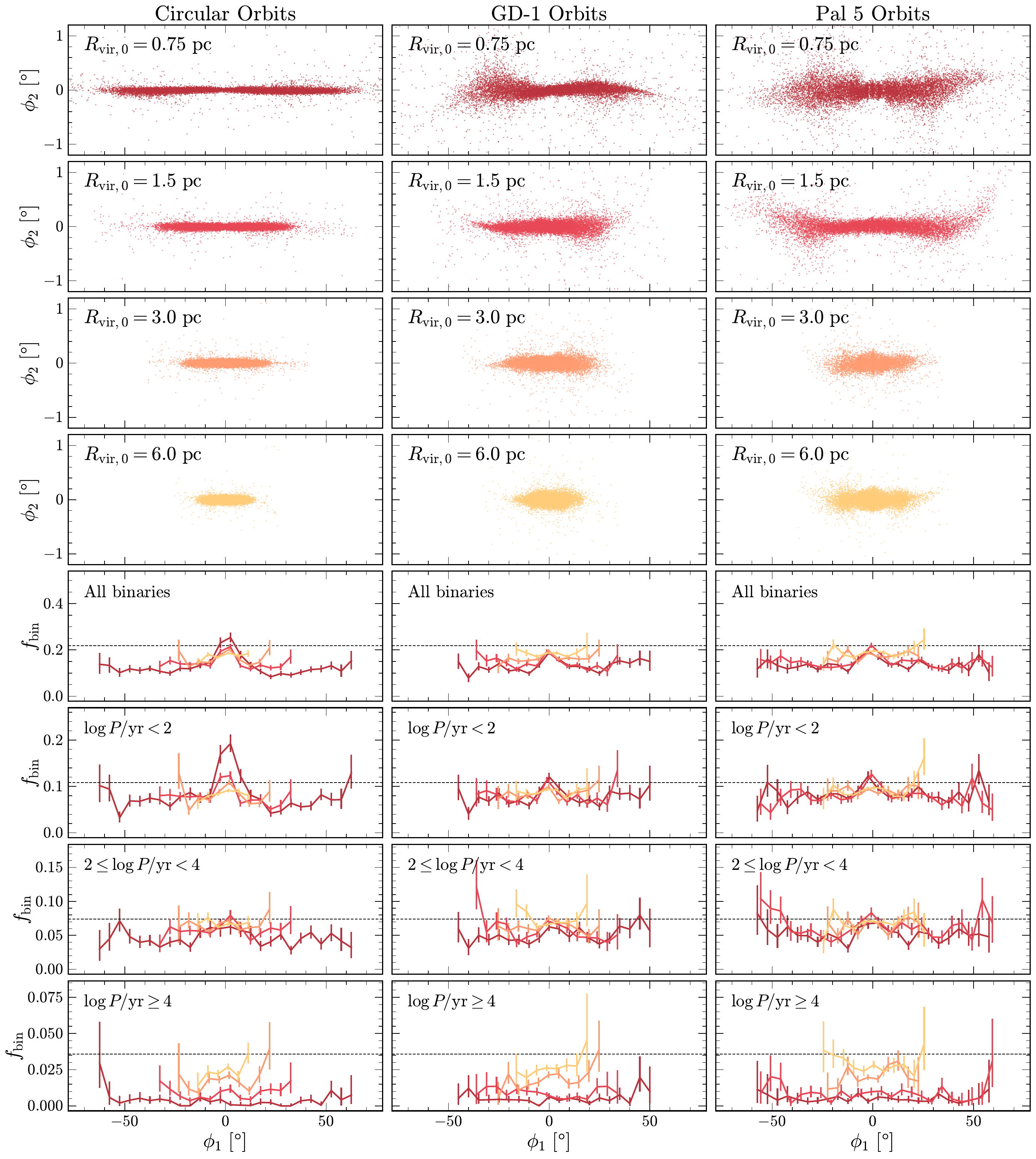}
    \caption{Same as Figure \ref{fig:fbin_lm}, but for the hi-OB simulations. Similarly, the colors corresponding to each simulation are the same as those in Figure \ref{fig:structural_evolution}, so that initially denser clusters are shown in darker shades.}
    \label{fig:fbin_hm}
\end{figure*}
\begin{figure*}
    \centering
    \includegraphics[width=\textwidth]{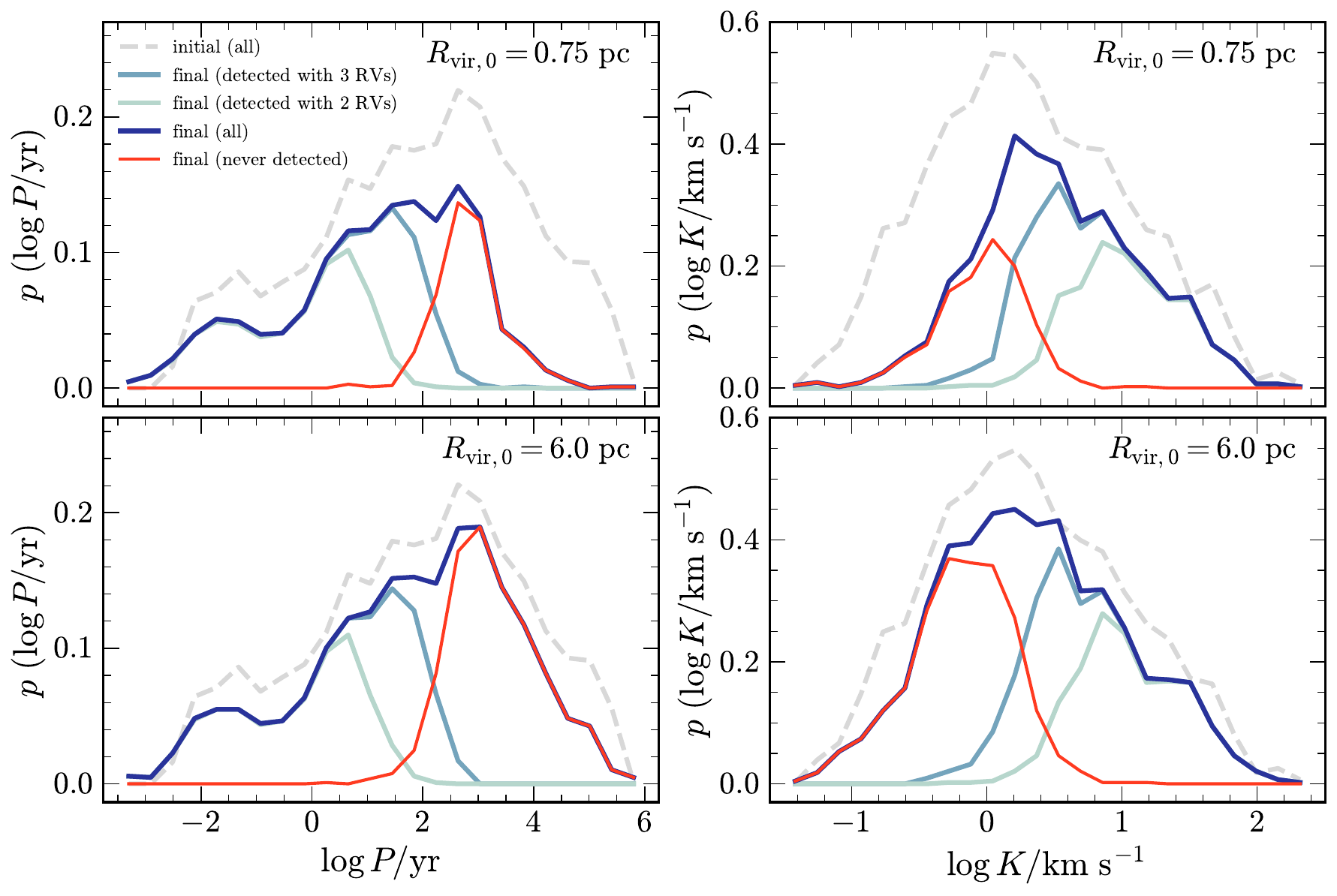}
    \caption{Period (left panels) and velocity semi-amplitude (right panels) distributions for the initial binary population (gray), the final binary population (dark blue), the binaries detected in a mock survey using two (light green) or three (light blue) RV measurements, and the binaries that are not detected in the mock survey (red). The simulations shown are the initially most dense (top panels) and least dense (bottom panels) lo-OB clusters on GD-1-like orbits. All histograms are expressed as probability densities normalized by the number of binaries in the initial population.}
    \label{fig:periods_outcomes}
\end{figure*}
\begin{figure*}
    \centering
    \includegraphics[width=\textwidth]{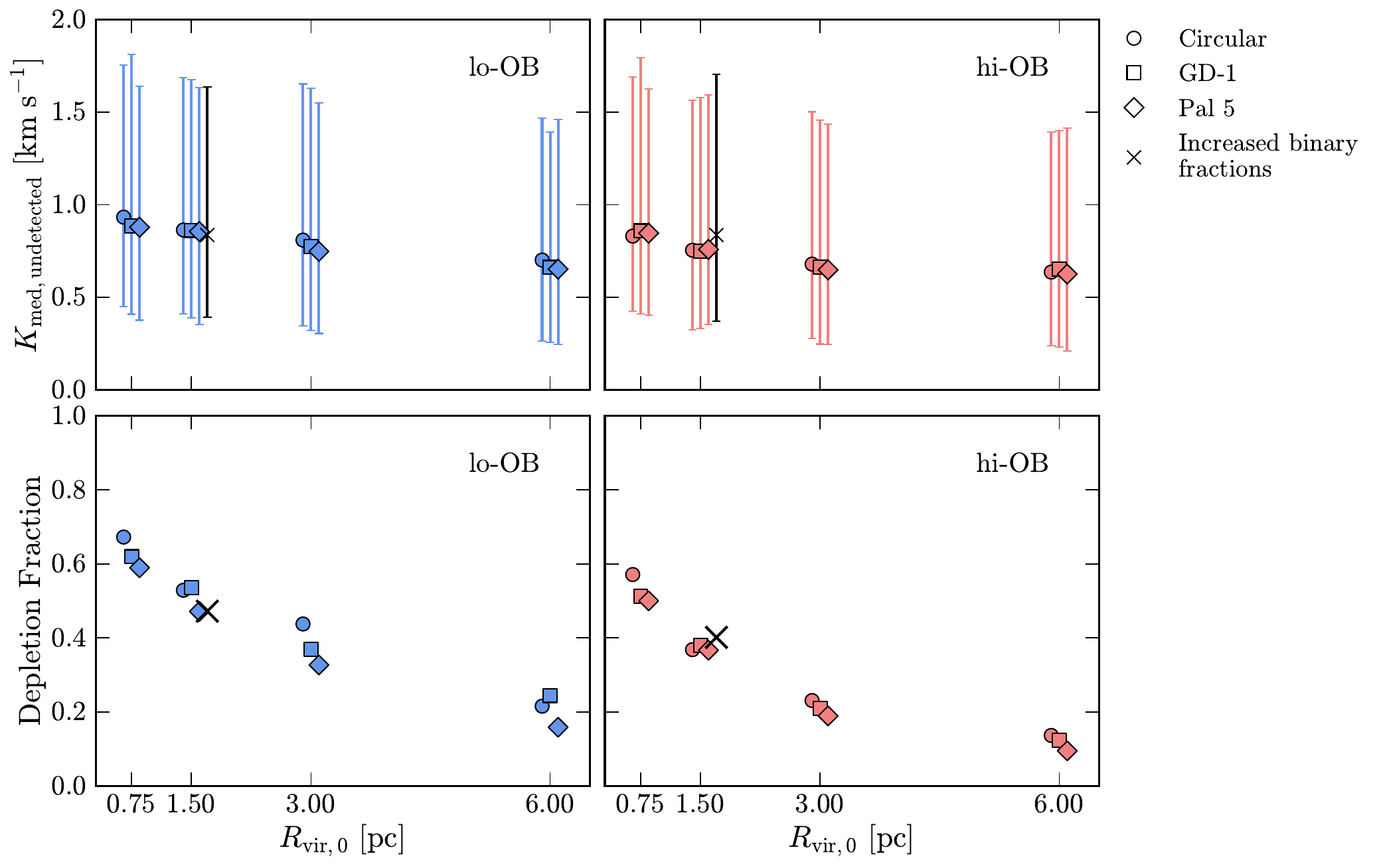}
    \caption{Summaries of the final undetected binary population in each simulation. In blue (pink) we show the lo-OB (hi-OB) streams from the main simulation grid, and in black, we show the clusters with altered binary populations. The stream with an overall doubled binary population is shown with the hi-OB simulations, and the stream with a doubled binary fraction only for $M_1<1.2~M_{\odot}$ is shown with the lo-OB simulations, as these initial $M_{\rm OB}$ values match most closely (see Table \ref{tab:simulation_grid}). \textbf{Top \textit{panels}}: Median velocity semi-amplitude of undetected binaries at the last pericenter passage before dissolution. We show the 16$^{\rm th}$--84$^{\rm th}$ percentile ranges in the data, and add small offsets to the initial virial radii to best display the error bars.
    \textbf{Bottom \textit{panels}}: The fraction of undetectable binaries depleted in each simulation. We perform a mock RV survey at $t=0$ and at the final pericenter passage of each cluster before dissolution to find the number of undetected binaries in each case, $N_{\mathrm{undetected}, i}$ and $N_{\mathrm{undetected}, f}$, respectively. The fraction displayed here is given by $1-N_{\mathrm{undetected}, f}/N_{\mathrm{undetected}, i}$. The same small offsets as in the top panels are added to the initial virial radii to best display all data points.}
    \label{fig:amp_detection_summary}
\end{figure*}
\begin{figure*}
    \centering
    \includegraphics[width=\linewidth]{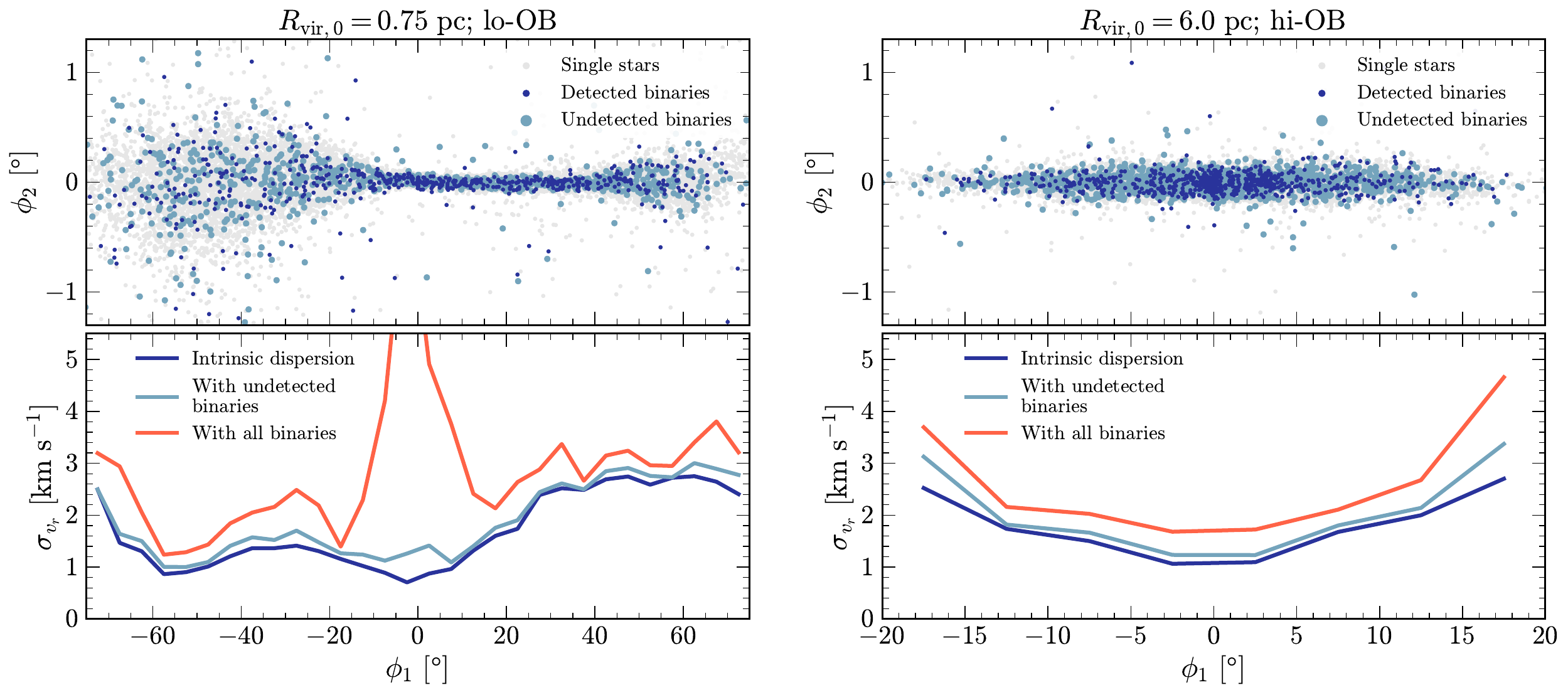}
    \caption{Locations of detected and undetected binaries in stream coordinates at the time of the last pericenter passage before dissolution (top panels) and radial velocity dispersion in $5\degree$ bins along $\phi_1$ at the same time using either the intrinsic velocities of systems in the stream, including the orbital velocities of the primary stars in undetected binaries, or including the orbital velocities of all primary stars in binaries, both detected and undetected (bottom panels). The simulations shown are the densest lo-OB cluster (left panels) and most diffuse hi-OB cluster (right panels) both on GD-1-like orbits. The overall velocity trend is removed by fitting and subtracting a 5th-order polynomial in $v_r$ vs $\phi_1$ before the dispersions are computed using the 5$^{\rm th}$--95$^{\rm th}$ percentile range in $\Delta v_r$ in each bin. As in Figures \ref{fig:fbin_lm} and \ref{fig:fbin_hm}, we require that $\geq50$ systems be present in a bin to include it in the bottom panels. }
    \label{fig:disp_along_stream}
\end{figure*}
Figures \ref{fig:fbin_lm} and \ref{fig:fbin_hm} display each simulated lo-OB and hi-OB stream, respectively, at the last pericenter passage of the cluster before its dissolution (see Table \ref{tab:simulation_grid}).
The top halves of the figures compare the stream morphologies, showing the streams in stream coordinates $\phi_2$ vs $\phi_1$. Clusters on circular orbits produce the thinnest streams, whereas the varying external potential over GD-1 and Pal 5-like orbits produces broader streams. The stream length at a given orbital phase increases with its age, so denser clusters with longer dissolution times have longer streams by the time they are shown. 

In the bottom halves of Figures \ref{fig:fbin_lm} and \ref{fig:fbin_hm}, we examine the binary populations binned in $\phi_1$ along the stream. We show the total binary fraction and the binary fraction in three period ranges: $P<100$ yr (``short periods"), $100\leq P<10^4$ yr (``intermediate periods"), and $P\geq10^4$ yr (``long periods"). The values of the binary fraction and their uncertainties are given by the median, $16^{\rm{th}}$, and $84^{\rm{th}}$ percentile values of the success fraction in the binomial distribution.

The binary fractions at short periods are centrally peaked, with the central maximum binary fraction often exceeding the initial value out to $|\phi_1|\sim10\degree$. The streams of denser and/or lo-OB clusters exhibit higher peaks in the short period binary fractions than the streams of less dense and/or hi-OB clusters.
At intermediate periods, the central peak in the binary fraction becomes less pronounced (typical for lo-OB streams) or disappears entirely (typical for hi-OB streams), and initially denser simulations tend to have lower binary fractions in the intermediate period range. 
At long periods, the binary fraction is roughly flat across all streams. Sometimes, the long-period binary fraction shows small increases at large $|\phi_1|$. They are the systems lost at early times and from large radii within the cluster, where they are less affected by two-body encounters prior to escape. 
The initially most diffuse simulations have the highest long-period binary fractions (depleted only to slightly below the initial values of $\sim3\%$), whereas the initially densest clusters have long-period binary fractions depleted nearly to zero.

\subsection{Mock RV survey}\label{sec:results4_mocksurvey}
We perform a mock RV survey of the stream data at the last pericenter passage of each cluster before their dissolution. We compute the velocity amplitudes of the binaries given their masses and semimajor axes, and generate their inclinations using a $\cos(i)$ isotropic distribution. We draw pericenter phases and arguments from uniform distributions $\mathcal{U}(0, 2\pi)$ and $\mathcal{U}(0,1)$ respectively, then compute the RV curve from Kepler's equations \citep{kepler1609}. We ``observe" the population of binaries for three epochs, with the second epoch 30--90 days after the first, and the third epoch 3--4 years after the second. 
We make no magnitude selection and sample the RV curve of every binary in the stream, including those containing stellar remnants. 
We assume a $0.1\ \rm{km\ s^{-1}}$ RV uncertainty and perform a chi-square test on the observed RVs with respect to their mean. We consider a binary to be ``detected" when the p-value of the chi square test is $<0.1$. When reporting the velocity semiamplitudes $K$ of binaries in the mock survey, we refer to the projected amplitude of the RV curve, accounting for the orbit's inclination. 

Figure \ref{fig:periods_outcomes} displays detailed results of the mock survey for the same simulations as in Figure \ref{fig:roche_hs} (the densest and most diffuse lo-OB streams on GD-1 orbits). We show the period and velocity semiamplitude distributions of the initial binary population, the final binary population across the entire stream, the binaries detected using the first two or all three RVs in the mock survey, and the binaries that remain undetected. 
The survey strategy allows for reliable detection of binaries with periods $\lesssim10^2$ yr (corresponding to $\sim1$--$10\ \rm km\ s^{-1}$ velocity semiamplitudes) regardless of the initial cluster density. On the other hand, wide binary depletion is dependent upon the initial cluster density, as shown in Figure \ref{fig:pDists}, so that the initially dense cluster disrupts most binaries with $P\gtrsim10^3$ yr ($K\lesssim0.5\ \rm km\ s^{-1}$), while the initially diffuse cluster leaves many of these binaries intact.

Figure \ref{fig:amp_detection_summary} summarizes the results of the mock survey for all simulations as a function of the initial virial radius. The top panels show the median velocity semiamplitude of the undetected binaries. Initially denser clusters produce streams in which the amplitudes of undetected binaries are on average higher ($\sim0.9\ \rm km\ s^{-1}$ when $R_{\rm vir,0}=0.75$ pc compared to $\sim0.6\ \rm km\ s^{-1}$ when $R_{\rm{vir,0}}=6$ pc). 
The bottom panels compare the number of undetected binaries at the final pericenter passage before dissolution to the number of undetected binaries when the mock survey is performed at the initial condition, yielding a ``depletion fraction" of undetected binaries $1 - N_{\mathrm{undetected}, f} / N_{\mathrm{undetected}, i}$.  Dynamics in initially denser clusters deplete higher fractions of the initially undetectable binaries (with $R_{\rm vir,0}=0.75$ pc clusters depleting $\sim 60\%$ of undetectable binaries compared to only $\sim10$--$20\%$ for the $R_{\rm{vir,0}}=6$ pc clusters).
Both the median semiamplitude of undetected binaries and undetectable binary depletion fraction are somewhat sensitive to the initial fraction of massive and short-lived stars in the cluster, with slightly higher median $K$ (by $\sim0.1~\rm km~s^{-1}$) and higher depletion fractions (by $\sim 10\%$) for the lo-OB clusters than for the hi-OB clusters of the same $R_{\rm vir,0}$. There is little difference in the median semiamplitude or depletion fraction across different orbit shapes.

Figure \ref{fig:amp_detection_summary} also includes the results for the simulations with altered binary fractions. We compare the simulation in which the binary fraction is doubled overall to the hi-OB streams and the simulation in which the binary fraction is doubled only below $M_1<1.2~M_{\odot}$ to the lo-OB streams, since this roughly matches their initial $M_{\rm OB}$ values (see Table \ref{tab:simulation_grid}). Both the typical undetected binary amplitudes and undetectable binary depletion fractions are consistent between the main simulation grid and the streams with altered binary populations of similar $M_{\rm OB}$.

We examine the consequences of including undetected binary orbits in measured RV dispersions along streams in more detail using a subset of the mock survey results in Figure \ref{fig:disp_along_stream}. The simulations that we show are the densest lo-OB cluster and the most diffuse hi-OB cluster, both on GD-1-like orbits. The locations of the detected and undetected binaries (using 3 RVs) in stream coordinates $\phi_1$ and $\phi_2$ are shown in the top panels, and the RV dispersions in bins along the stream longitude are shown in the bottom panels. We show both the intrinsic dispersions (i.e., using the velocities of the single stars and binary centers of mass), and the dispersion including the orbital velocities of the primary in each undetected binary. We also show the velocity dispersions including the primary star orbital motions for all binaries, both detected and undetected. The overarching velocity trend was removed prior to computing the dispersions by fitting and subtracting a 5th-order polynomial to the stream in $v_r$ vs $\phi_1$, similarly to our straightening of the streams in $\phi_2$ vs $\phi_1$. The reported dispersions are the 5$^{\rm th}$--95$^{\rm th}$ percentile range, which we choose over the standard deviation because it is more robust to outliers. 

Despite the fact that typical RV amplitudes of undetected binaries are of order $1~\rm km~s^{-1}$, the net dispersion added to the streams is much lower, only of order $\sim0.1~\rm km~s^{-1}$. The amount of added dispersion from undetected binaries is relatively uniform across both streams, though there is a slight increase in the added dispersion near the progenitor lo-OB dense cluster which is not present for the hi-OB diffuse cluster. This is because the dense lo-OB cluster has undergone significant mass segregation, causing a centrally-peaked binary fraction in its stream (see Figure \ref{fig:fbin_lm}), while the diffuse hi-OB cluster has not. This effect is more apparent when comparing the added dispersion from all binaries. For the diffuse hi-OB cluster with little mass/binary segregation, short-period, high-amplitude binaries are spread evenly along the stream so that they add a uniform $\sim1~\rm km~s^{-1}$ to the {5$^{\rm th}$--95$^{\rm th}$ percentile range of RVs in each bin compared to the intrinsic value. For the mass segregated, dense/lo-OB cluster, the over-abundance of high-amplitude binaries near $|\phi_1|=0$ increases the dispersion significantly, by $\gtrsim4~\rm km~s^{-1}$ where $|\phi_1|\lesssim10\degree$. 
There are also slight increases in the added dispersion at large $|\phi_1|$, in particular for the diffuse and hi-OB cluster, which may be due to decreasing numbers of systems in the outer bins. 
Finally, the intrinsic dispersions themselves show rich morphologies which vary with the progenitor cluster properties and orbit. A detailed description of the intrinsic dynamical structures of streams and their measurability given the stream binary population is left to future work.

\section{Discussion}\label{sec:discussion}
\subsection{Dynamically shaping the stream binary population}\label{sec:discussion_1_clusterprocessing}
Here, we provide a summary of the evolution of the stream binary population (see Figures \ref{fig:structural_evolution}--\ref{fig:fbin_hm}). 

The widest binaries (with $P>P_{\rm crit}$) are disrupted rapidly by tides internal to the cluster, regardless of interactions with other stars. The number of systems disrupted in this manner increases with the cluster initial density (Figures 
\ref{fig:pDists}, \ref{fig:roche_hs}). The cluster density drops over a timescale of order $t_{\rm int\ tid}=t_{\rm SE}(M\sim 25~M_{\odot})+t_{\rm dyn}$ as it expands due to mass loss from the evolution and deaths of its most massive stars. After the cluster expands, internal tides become a less important mechanism for binary disruption. 
The remaining soft binaries in the cluster are disrupted over a relaxation timescale through two-body encounters. 

Meanwhile, the cluster exhibits mass segregation to a degree dependent on its initial density and fraction of massive and short-lived stars. High initial densities facilitate rapid mass segregation (top panels of Figure \ref{fig:bound_MFs}), while lower initial densities lead to slower mass segregation (bottom panels of Figure \ref{fig:bound_MFs}). Cluster expansion due to mass loss from stellar evolution drives up the cluster relaxation timescale and shortens its lifetime, allowing for less mass segregation (bottom right panel of Figure \ref{fig:bound_MFs}).

The approximate order in which stars escape a cluster is preserved in their ordering along the tails of the resulting stellar stream, so that stars that escape the cluster earlier are located at larger $|\phi_1|$ (e.g., \citealt{webbVariationStellarMass2021}). As a result, the combined effects of binary disruption (removing soft binaries while they remain bound to the cluster) and mass segregation (acting to keep surviving binaries bound to the cluster while single stars preferentially escape) produce a period-dependent spatial distribution of binaries along the stream. 
Soft binaries must escape dense cluster environments at early times in order to survive in the simulation until $t_{\rm dis}$. These systems are present only at large $|\phi_1|$ in the resulting streams. This result is similar to findings from \citet{Wang2024}, in which the binary fraction in \textsc{petar} simulations increases at large radial distances from the progenitor cluster due to early escapers. 
Wide binaries may also be present in the streams of initially diffuse clusters, which do not disrupt binaries with internal tides and have large hard/soft boundaries.
On the other hand, the spatial distribution of close binaries along the stream is dictated by mass segregation. Hard binaries escape mass segregated clusters at late times, leading to centrally peaked close binary fractions along the stream. Centrally peaked close binary fractions are absent from the streams of diffuse clusters with higher fractions of massive and short-lived stars because they undergo less mass segregation.

\subsection{Implications for dark matter searches}\label{sec:discussion_3_DM}

The dynamical state of stellar streams is sensitive to perturbations by dark matter subhalos, either individually in the form of large-amplitude (0.1--1 km s$^{-1}$) perturbations or collectively in the form of a net dynamical heating. In the case of individual perturbations, RV variations from binaries will present a source of noise, and in the case of accumulated dynamical heating, these variations will present a source of bias. 
We have shown that a strategically-chosen RV measurement cadence over $\sim5$ yr should allow for the detection of most binaries with orbital periods $\lesssim100$ yr, and that undetectable binaries are dynamically depleted within the cluster, so that only $\sim10$--$60\%$ of undetectable binaries initially present in the cluster remain in the stream (Figure \ref{fig:amp_detection_summary}). 
Binaries with periods $100\lesssim P\lesssim10^3$--$10^5$ yr (depending on the initial cluster density) both survive dynamical processing within the cluster and evade detection in the stream. They have typical velocity amplitudes of $0.5\lesssim K\lesssim1\ \rm km\ s^{-1}$, and increase measured stream velocity dispersions by $\sim0.1~\rm km~s^{-1}$ (Figures \ref{fig:amp_detection_summary} and \ref{fig:disp_along_stream}). We have also underscored the importance of multi-epoch observations---the added dispersion from all binaries in a single RV epoch is as high as $>5~\rm km~s^{-1}$ (Figure \ref{fig:disp_along_stream}).

While the fractional depletion of undetectable binaries is insensitive to the primordial binary fraction (Figure \ref{fig:amp_detection_summary}), the abundance of undetectable binaries present in the stream still reflects the initial binary demographics of the progenitor. Uncertainties in the primordial binary fraction of the stream progenitor (see Section \ref{sec:discussion_4_limitations}) therefore place fundamental limitations on our understanding of the present day stream binary population. 
The consequences are twofold. 
First, the net velocity dispersion increase to streams from undetected binaries is not robust to the initial binary fraction. Constraints on subhalo populations based on stream heating cannot safely assume a fixed binary-induced dispersion of $0.1~\rm km~s^{-1}$, but rather must allow for the possibility that undetected binaries with amplitudes of order $1~\rm km\ s^{-1}$ make up an unknown fraction of the stream stars. 
Second, the amount of noise added to characteristic velocity signatures of stream/subhalo encounters by undetected binaries cannot be predicted without knowledge of the initial cluster binary fraction. However, because binaries present a source of noise rather than bias to measurements of individual subhalo encounters, it is possible that these perturbations will still be observable and provide valid dark matter constraints given a sufficient number of observed targets. 
A thorough exploration of the implications of our results for the variety of methods to constrain the nature of dark matter using stream kinematics is left to future work.

\subsection{Caveats \& Limitations}\label{sec:discussion_4_limitations}

\emph{Cluster initial conditions}: We have initialized clusters as if their stars are born distributed as \citet{king1966} profiles on halo-like orbits. However, stars may form and begin to evolve during cloud collapse. In this scenario, stellar winds dissipate residual gas in the protocluster (e.g., \citealt{hills1975}) before violent relaxation can fully drive the system toward a King distribution function. Further, the dynamical ages implied by the observed lengths of present-day MW stellar streams are typically younger than their stellar ages, reflecting cluster formation sites different from their present-day orbits (e.g., due to having been accreted onto the MW halo from a lower mass satellite rather than forming in situ; \citealt{helmi1999, panithanpaisalBreaking$textsfCosmoGEMS$Modeling2025}).
Self-consistent treatments of clusters evolved within cosmological simulations have been explored in recent work (see \citealt{rodriguezGreatBallsFIRE2023, panithanpaisalBreaking$textsfCosmoGEMS$Modeling2025}), though these rely on Monte Carlo techniques rather than direct $N$-body calculations to model cluster evolution and the demographics and escape conditions of the stream stars.

Our simulation grid explores only clusters of $N=15000$ stars. 
This reflects the lowest-mass minority of present-day GCs in the halo (of interest to dark matter searches due to the small intrinsic velocity dispersions of their streams), but not the originally formed Milky Way GC population. The initial masses of clusters on Pal~5-like orbits are especially unlikely to reflect the actual Pal~5 progenitor. The Pal~5 stream is $\gtrsim 3$ Gyr old, but maintains a surviving progenitor in contrast to the shorter dissolution times of typical clusters on Pal~5-like orbits in this work (e.g., \citealt{bonacaVariationsWidthDensity2020,Gieles2021, Wang2024}). 
For clusters with larger $N$, the stochastic variation in the fraction of massive and short-lived stars would be suppressed and such clusters would show even less variation in wide binary depletion than we observe here between the lo-OB and hi-OB clusters. 
We also explore only a limited range of cluster orbit shapes, and note that future work will tune both $N$ and the cluster orbit to create realistic mocks of MW streams at present day. 
The disruption of wide binaries by internal tides at early times implies a sensitivity to initial conditions such as primordial mass segregation and cluster concentration, which we do not explore here. 

\emph{Binary modeling}: The binary demographics of low-mass primaries in our initial condition are extrapolated, rather than observationally constrained, and our main simulation grid likely underestimates the primordial binary fraction for stars $<0.8\ M_{\odot}$ \citep{wintersSolarNeighborhoodXLV2019, offnerOriginEvolutionMultiple2023}. 
While we show that the fraction of undetectable binaries depleted by cluster dynamics is insensitive to the initial low-mass binary fraction, the absolute binary fraction of M-dwarfs at the end of the simulation is certainly sensitive to its initial value. 
An increase in the M-dwarf binary fraction may allow for more M-dwarf binaries to mass segregate, displacing higher mass (and more readily observable) single stars into the stream. 

Even in the mass range where our binary statistics are constrained observationally by \citet{moe2017}, the constraining observations are of solar-metallicity binaries. Since the close binary fraction is anticorrelated with metallicity \citep{moeCloseBinaryFraction2019}, the primordial binary populations of metal-poor GCs should have contained a higher fraction of close binaries than what we model here. This may represent a meaningful limitation, as massive close binaries play an important role in GC dynamics and in particular the degree of mass segregation experienced by surviving binaries. 

Finally, the binary evolution physics in the BSE package is simplified and approximate. While computationally efficient, it lacks the self-consistent modeling of binary mass transfer, a proper accounting for angular momentum transport between binary components, and the accuracy of the final compact object population compared to more detailed binary population synthesis models (e.g., \citealt{fragosPOSYDONGeneralpurposePopulation2023}). 
Changes to the binary evolution physics (and/or to the demographics of massive, close, and interacting binaries) would likely influence the global cluster structural evolution over multiple relaxation timescales by altering the population of compact objects present in the cluster. For example, \citet{Aros2021} show that the presence of intermediate-mass black holes in the centers of GCs both halt the segregation of binaries toward the cluster core and disrupt the binaries that do segregate. 
However, we have shown that the fraction of wide and undetectable binaries disrupted within the cluster and the typical RV amplitudes of remaining undetectable binaries are more sensitive to the initial cluster density than to the stellar population or the initial binary demographics.

\section{Summary and Future work}\label{sec:summary}
We examined the behavior of a realistic population of binaries in direct $N$-body simulations of GCs and their resulting stellar streams, finding that the final distribution of binaries along the stream is period dependent and sensitive to the initial cluster density and fraction of massive and short-lived stars. 

Wide binaries ($P\gtrsim10^3$--$10^5~\rm yr$) are destroyed by internal tides before cluster expansion due to massive stellar evolution (over several dynamical times), and by two-body encounters over the cluster relaxation time. Initially denser clusters dynamically deplete more binaries and create streams with lower wide binary fractions. 
Close binaries ($P\lesssim10^{2}~\rm yr$) preferentially remain in the cluster due to mass segregation, creating stream binary fractions that peak at $\phi_1=0\degree$. High cluster densities and relatively few OB stars result in higher central peaks in the close binary fraction along the stream. 
In mock RV surveys, undetectable binaries have RV amplitudes of $\sim0.5$--$1~\rm km~s^{-1}$ and add $\sim0.1~\rm km\ s^{-1}$ of RV dispersion to the simulated streams, after being depleted by $\sim10$--$60\%$ through cluster dynamics. 

Our grid of models represents the first step in a direct $N$-body approach to a more complete understanding of the dynamical structures of GC streams. Future work will improve the realism of our mock observations in order to best characterize the level of dispersion added by undetected binaries. The intrinsic stream dispersions and the dispersions from undetected binaries will then be compared to the amplitudes of individual perturbations from dark matter subhalos, and the dispersion added by accumulated impacts of subhalo populations in different cosmologies. 
Multi-epoch precision RV measurements of Milky Way streams (e.g., in the upcoming Via Survey\footnote{\href{https://via-project.org/}{https://via-project.org/}}) will reveal their true spatial distribution of close binaries and test the ability of our models to predict the full binary demographics on the route to providing dark matter constraints using stream kinematics. 

\begin{acknowledgements}
We thank Kareem El-Badry for his modification of the \texttt{COSMIC} binary population to produce the correct initial mass function, Phoebe Heretz for suggestions to improve early drafts of the manuscript, and Newlin Weatherford for helpful discussions related to cluster dynamics. 
AP acknowledges support from the National Science Foundation Graduate Research Fellowship under Grant No. DGE 2140743. 
JN is supported by a National Science Foundation Graduate Research
Fellowship, Grant No. DGE-2039656. 
LW thanks the support from the National Natural Science Foundation of China through grant 12573041 and 12233013, the High-level Youth Talent Project (Provincial Financial Allocation) through the grant 2023HYSPT0706, the Fundamental Research Funds for the Central Universities, Sun Yat-sen University (2025QNPY04).
\end{acknowledgements}

\bibliography{ref}{}
\bibliographystyle{aasjournal}



\end{document}